\documentclass[nohyper]{JHEP3}

\usepackage[dvips]{graphicx}

\renewcommand{\thefigure}{\Roman{figure}}
\def\beq{\begin{equation}}
\def\eeq{\end{equation}}
\def\bea{\begin{eqnarray}}
\def\eea{\end{eqnarray}}
\def\bq{\begin{quote}}
\def\eq{\end{quote}}
\def \lsim{\mathrel{\vcenter
     {\hbox{$<$}\nointerlineskip\hbox{$\sim$}}}}
\def \gsim{\mathrel{\vcenter
     {\hbox{$>$}\nointerlineskip\hbox{$\sim$}}}}
\def\gappeq{\mathrel{\rlap {\raise.5ex\hbox{$>$}}
{\lower.5ex\hbox{$\sim$}}}}
\def\lappeq{\mathrel{\rlap{\raise.5ex\hbox{$<$}}
{\lower.5ex\hbox{$\sim$}}}}

\def\bea{\begin{eqnarray}}   
\def\eea{\end{eqnarray}}

\title{CP Violation in the SUSY Seesaw:\\ Leptogenesis
 and Low Energy}

\author{S. Davidson\\
IPN de Lyon, Universit\'e Lyon 1, CNRS \\ 
 4 rue Enrico Fermi, Villeurbanne,  69622 cedex France\\
 E-mail: \email{s.davidson@ipnl.in2p3.fr}}
\author{J. Garayoa\\
Depto.\ de F\'{\i}sica Te\'orica,
and IFIC, Universidad de
Valencia-CSIC \\ 
Edificio de Institutos de Paterna, Apt. 22085, 46071 Valencia,
Spain\\
E-mail: \email{garayoa@ific.uv.es} }
\author{F. Palorini\\
IPN de Lyon, Universit\'e Lyon 1,  \\ 
 4 rue Enrico Fermi, Villeurbanne,  69622 cedex France\\
E-mail: \email{f.palorini@ipnl.in2p3.fr}}
\author{N. Rius\\
Depto.\ de F\'{\i}sica Te\'orica and
IFIC, Universidad de
Valencia-CSIC \\ 
Edificio de Institutos de Paterna, Apt. 22085, 46071 Valencia,
Spain\\
E-mail: \email{nuria@ific.uv.es} }

\abstract{ We suppose that the baryon asymmetry is produced by
 thermal leptogenesis (with flavour effects), at  temperatures
 $\sim 10^{9} - 10^{10}$ GeV,  in the supersymmetric seesaw
 with universal and  real soft terms. The parameter space
 is restricted by assuming that $\ell_\alpha
 \to \ell_\beta \gamma$ processes will be
 seen in upcoming experiments. We study
 the sensitivity of the baryon asymmetry to the
 phases  of the lepton mixing matrix,
 and find that leptogenesis can work for any
 value of the phases.
We also estimate  the contribution to the 
electric dipole moment of the electron,
arising from  the seesaw,  
and  find that 
it is (just) beyond  the sensitivity of next generation experiments
($\lsim 10^{-29} e$ cm). 
The fourteen  dimensional  parameter space is
efficiently  explored with a Monte Carlo Markov Chain, which
concentrates on the regions of interest.
}

\preprint{LYCEN 2008-06\\
IFIC/08-28\\
FTUV-08-0419}

\begin{document}

\section{Introduction}
\label{Introduction}

Neutrino masses are evidence for  
beyond the Standard Model (SM) physics.
A simple extension of the standard model that accounts for neutrino 
masses is the seesaw mechanism
\cite{ss}, where heavy majorana right-handed neutrinos are added to the SM. 
Moreover, 
the seesaw scenario provides a very attractive framework to explain the baryon 
asymmetry of the universe (BAU) through the leptogenesis \cite{fy}  mechanism,
without inducing  proton decay.

CP violation is a necessary ingredient to explain the BAU and, if 
this asymmetry is produced via leptogenesis, the required CP violation  is 
encoded in the CP violating phases of the lepton sector. 
Three of them are
the well known Dirac and Majorana phases of the PMNS mixing matrix, that 
are in principle  
 measurable. Any observation of CP violation 
in the lepton sector, for instance CP violation in neutrino oscillations due 
to the PMNS phase $\delta$, would then support leptogenesis
by demonstrating that CP is not a symmetry of leptons. However, even in this 
very promising case, the question  of whether  the BAU is
produced via leptogenesis  is far 
from  being answered, because it is not possible to reconstruct  the 
high-energy CP odd observables from the low-energy ones 
\cite{Davidson:2001zk} without assuming very 
constraining frameworks for the unmeasurable quantities. 
Therefore, the intent of this work is  to clarify the relation between the CP 
violation accessible to low-energy experiments,
 and the CP violation necessary 
for leptogenesis,  in a  phenomenological {\it bottom-up} perspective, 
with minimal assumptions about the high scale theory.
We just assume that the neutrino Yukawa 
couplings are hierarchical, 
which is the most natural assumption 
given the observed values of the charged lepton and quark Yukawas.
Neutrino oscillation data then lead to 
hierarchical singlet masses.

In this paper, we aim
to answer the phenomenological question of whether the BAU can be 
{\it sensitive}  to low-energy phases, in the supersymmetric
seesaw. We suppose the observed BAU is generated via thermal leptogenesis,
and enquire whether this restricts the  range of the phases.  
A similar issue was investigated by Branco {\it et.al}  \cite{Branco:2001pq}, 
where it was shown that for any value of the measurable CP violating phases, 
a large enough BAU can be produced. This statement has been recently 
confirmed in a study \cite{Davidson:2007va} that includes flavour effects \cite{flavour}, in the 
Standard Model seesaw framework.
In the present analysis, we want to address the question considering 
flavoured leptogenesis in a supersymmetric scenario, that has the interesting 
feature to potentially add new observables in the lepton sector, through the 
enhancement of flavour and CP violating processes 
(See {\it eg} \cite{Raidal:2008jk} for a review and references on
leptonic flavour and CP violation, induced by supersymmetry.).

The question we address, and the answer we find, differ from
some other analyses \cite{Pascoli:2006ie,Pascoli:2006ci,Anisimov:2007mw,Molinaro:2008rg}. 
As written above,
we aim to make few untestable assumptions, and to ask a precise
phenomenological question: ``Is the baryon asymmetry sensitive
to PMNS phases?''. We find the answer to be no.  That is, there
is { ``no correlation''}  between the BAU and PMNS phases, when
all the unmeasurables in our scenario are allowed to vary
over their whole range. To the best of our
understanding, Refs.  \cite{Pascoli:2006ie,Pascoli:2006ci,Anisimov:2007mw,Molinaro:2008rg} 
find a correlation between
the BAU and the PMNS phases because they set
unmeasurables (such as phases of the ``right-handed''
neutrinos) to fixed values.

 We define ``finding a correlation between $Y_B$ and $x$'' to mean
``$Y_B$ is sensitive to $x$''.  To show  that  the baryon
asymmetry $Y_B$ is insensitive to (or uncorrelated with)
a parameter  $x$,
we must only show that, for any value of $x$, we
can find a large enough $Y_B$.
It would be numerically more challenging to show   a correlation,
because the point distribution in scatter
plots may reflect the priors on the scanned parameters  (see
sections \ref{convergence}
and \ref{method}).
Our definition of correlation
differs from that used by \cite{Pascoli:2006ie,Pascoli:2006ci,Molinaro:2008rg},
and also   in \cite{Ellis:2002xg} (who extract correlations from scatter plots).
 We use our narrow definition because it is parametrisation
independent.

Since leptogenesis occurs at a very high-energy scale, a supersymmetric 
scenario is desirable in order to stabilize the hierarchy between the 
leptogenesis scale and the electroweak one. However, 
if supersymmetry exists at all, it must be broken and, in principle, the soft 
supersymmetry breaking Lagrangian can contain off-diagonal (in flavour space) 
soft terms, that would enhance lepton flavour violating (LFV) processes.
These are strongly constrained by current 
experiments; this is the so-called supersymmetric flavour problem. 
In order to avoid it, we focus on the most conservative minimal Supergravity 
(MSUGRA) scenario with real boundary conditions, 
where the dynamics responsible for supersymmetry breaking are flavour blind 
and all the lepton flavour and CP 
violation is controlled by the neutrino Yukawa couplings.
Supersymmetric expectations for LFV \cite{Borzumati:1986qx,Hisano:1995cp,meg}
and possible relations
to leptogenesis 
\cite{Raidal:2008jk,Molinaro:2008rg,AT,Branco:2002kt,afs,Petcov:2006pc}
{ \footnote{See ref. \cite{afs} for a discussion about when 
the approximation used in \cite{Branco:2002kt} is not valid.}}  
and EDMs 
\cite{Ellis:2002xg,JMR} have been studied by many people.

We perform a scan over the seesaw parameters, looking for 
those points that give a large enough BAU, and 
where $\mu \to e \gamma$ and one of $\tau \to \ell \gamma$ would be
seen in upcoming experiments. { Our analysis
is more restrictive than \cite{Ellis:2002xg}, in that we
require these branching ratios to be ``large''.} 
The aim is to verify if such experimental inputs imply a preferred range of 
values for the low-energy PMNS phases. We also estimate
 the contribution to 
the CP violating electron  electric dipole moment.
A detailed analysis of the MSUGRA scenario 
would require a scan also over the supersymmetric parameters, which is 
beyond the scope of our analysis.

Due to the large number of unknown parameters, instead of doing a usual grid 
scan in the seesaw parameter space we construct a Markov Chain using a 
Monte Carlo simulation (MCMC --- see {\it e.g.} \cite{MacKay,Gilks}). 
This technique
allows to efficiently explore a high-dimension parameter space, and we
apply it for the first time to the supersymmetric seesaw model
\footnote{See \cite{nops} for a detailed study of the Zee-Babu model 
of neutrino masses phenomenology using this technique.}.
Our work is thus pioneering in the exhaustive scanning of the seesaw
parameters, which would be otherwise prohibitive without the MCMC
technique.

The paper is organized as follows. In section \ref{Notation} we introduce 
the supersymmetric seesaw in the MSUGRA scenario and  
we review the low-energy  interactions induced in the
supersymmetric seesaw model.
Section \ref{leptogenesis} is devoted to thermal leptogenesis
with flavour effects, and 
 section \ref{Reconstructing} describes our bottom-up 
reconstruction procedure.
Section \ref{Analytic} gives  analytic estimates, that complement
 our numerical  analysis, using the  MCMC technique, 
which is 
presented in  section \ref{MCMC}.
 We discuss our results in
section \ref{Discuss} and conclude in   section 
 \ref{Conclusions}.

\section{Notation and review}
\label{Notation}

We consider the superpotential for the leptonic sector in a 
supersymmetric seesaw model \cite{ss}
with three hierarchical right-handed neutrinos ($M_1<M_2<M_3$):
\beq
W_{lep} =
 (L_L H_d)  { Y_e}  E^c +
 ({L_L}  H_u) { \lambda} N^c +  N^c\frac{M}{2} N^c.
\label{W}
\eeq 
In this expression, $\lambda, Y_e$ and $M$ are $3 \times 3$ matrices,
and flavour indices are suppressed.
The  $L_L$ are the supermultiplets containing left-handed 
lepton fields, 
$E$ are those containing the right-handed charged leptons, while $N$ are the 
supermultiplets of the right-handed singlets. The Majorana mass scale can 
be taken large $10^9$ GeV $  \lsim M_i \lsim 10^{15}$ GeV, 
since the corresponding operator is a singlet under 
the SM gauge group.

Without loss of generality one can work in the basis where $Y_e$ and $M$ 
are diagonal, so that the superpotential gives the following Lagrangian for 
leptons:
\beq
{\cal L} =   Y_{e_{\alpha}} (\overline{\ell}_L^{\alpha}  H^*_d)  e_R^{\alpha}+ 
   (\overline{\ell}_L^{\alpha}   H^*_u)  {\bf  \lambda^*}_{\alpha i} N_i  + 
 \frac{{M}^i}{2}  \overline{N^c}_i N_i +...+ h.c.
\eeq
where the parentheses indicate SU(2) contractions and 
the flavour indices are written explicitly. Since 
supersymmetry is broken, to this Lagrangian we must add the soft SUSY 
breaking terms :  
\beq
\mathcal{L}_{SSB} =  \tilde{m}_0^2  \sum_f 
\tilde{f}^{\dagger} \tilde{f} +
\left\{ \frac{B M^i}{2} \tilde{N}_i^c \tilde{N}_i^c
+ a_0( y_{e_{\alpha}} {\tilde{\ell}_L}^{\alpha} \cdot  H_d  \tilde{e}_{\alpha}^c +
\lambda_{\alpha i} \tilde{\ell}^{\alpha}_L \cdot  H_u \tilde{N}_i^c)  + h.c. \right\}
\label{L}
\eeq
where $\tilde{f}$ collectively represents sfermions.
This soft part is written at some high scale $M_X$ where, in MSUGRA, 
the soft masses are 
universal and the trilinear couplings are proportional to the corresponding 
Yukawas.
MSUGRA is then characterized by four parameters: 
the scalar ($m_0$) and gaugino ($m_{1/2}$) masses, 
shared by all of them at the GUT scale; 
the trilinear coupling involving scalars, $a_0$, at the 
GUT scale; and finally the Higgs vev ratio, $\tan \beta$.

 In the chosen basis, the neutrino Yukawa matrix is in general not 
diagonal and complex, and can be written as:
\beq \label{yuk}
\lambda = V_L^{\dagger} D_{\lambda} V_R
\eeq
where $D_{\lambda}$ is diagonal and real. 
Note that in this basis the neutrino Yukawa matrix is the only 
source of flavour violation in the lepton sector, through the unitary matrices 
$V_L$ and $V_R$ that act respectively on the lepton doublet space and on the 
right-handed neutrino space. These matrices contribute also to CP violation, 
through six CP violating phases. In general, other sources of CP violation 
appear in the complex neutrino B-term, in the scalar mass $\tilde{m}_0$ and in 
the trilinear coupling $a_0$.

At energies well below the right-handed neutrino mass scale, 
the effective light neutrino
majorana mass matrix can be written:
\beq \label{m_nu_eff}
[m_{\nu}] = \lambda M^{-1} \lambda^{T} v_u^2 = U D_{\nu} U^T.
\eeq  
The first equality shows that the smallness of light neutrino masses is 
naturally explained once the right-handed neutrino mass is set at very high 
energy, $\sim 10^{14}$ GeV (in this expression $v_u = \langle H_u \rangle$). 
In the second equality, $D_{\nu}$ is a diagonal matrix with real 
positive eigenvalues and $U$ is the $PMNS$ matrix containing the three  
low-energy CP violating phases, the Dirac phase $\delta$ and two Majorana
phases $\alpha$, $\beta$. Those phases are, in general, a combination of the 
6 phases appearing in the complete theory.
We use the standard parametrisation:
\begin{equation} \label{U_PMNS}
    U =\left(
    \begin{array}{ccc}
         e^{i \alpha}~ c_{13} c_{12}
        &e^{i \beta}~ s_{12} c_{13}
        & s_{13}~ e^{-i \delta}  \\
         e^{i \alpha}~(-s_{12} c_{23} - s_{23} s_{13} c_{12}~e^{i \delta})  
        &  e^{i \beta}~(c_{23} c_{12} - s_{23} s_{13} s_{12}~ e^{i \delta}) 
        &  s_{23} c_{13} \\
        e^{i \alpha}~ (s_{23} s_{12} - s_{13} c_{23} c_{12}~e^{i \delta})
       &  e^{i \beta}~(-s_{23} c_{12} - s_{13} s_{12} c_{23}~e^{i \delta})
        &  c_{23} c_{13}
    \end{array} \right) .
\end{equation}
If we combine the equations (\ref{yuk}) and (\ref{m_nu_eff}), we can write:
\beq \label{Dnu}
D_\nu  = U^\dagger V_L^\dagger D_\lambda V_R D_M^{-1}V_R^T D_\lambda
V_L^* U^* v_u^2 \equiv W^\dagger D_\lambda V_R D_M^{-1}V_R^T D_\lambda
W^* v_u^2 ,
\eeq
with $V_R$ diagonalizing the inverted right-handed
neutrino mass matrix. This relation shows that non-zero angles and phases 
in the unmeasurable right-handed neutrino mixing matrix $V_R$
imply non-zero angles and phases in $W = V_L U$, which being
in the doublet sector, is potentially more accessible.
We will use this relation to reconstruct the right-handed sector 
from low energy physics in sec. \ref{Reconstructing}.

\subsection{Low-energy footprints: LFV and EDMs in MSUGRA}
\label{MSUGRA}

Present bounds on LFV processes, shown in table 
\ref{table:LFVrates},  restrict
the size of flavour off-diagonal 
soft terms. This  suggests    universal soft terms
  at some high 
scale $M_X$, see Eq. (\ref{L}), like in the MSUGRA scenario. 
There are also stringent 
experimental bounds, as we can see in Table (\ref{table:EDMbounds}), 
on the CP violating electric dipole moments, which point towards very
small CP phases. To address this ``SUSY CP problem'' \footnote{See
{\it e.g.} \cite{oscar} for an illuminating discussion.}, 
we  suppose that all the soft 
breaking terms (namely $a_0$, $m_0$ and right-handed sneutrino B-term),
as well as the $\mu$ term, are real.
Even under this extremely conservative assumptions, it is well known that 
because of RGE running 
from high to low energy scales, the seesaw Yukawa couplings potentially
induce lepton flavour and CP violating contributions to the soft terms 
\cite{Borzumati:1986qx,Hisano:1995cp,meg}. 

\begin{table}
\begin{center}
\begin{tabular}{|c|c|c|}
\hline
& Present bounds & Future sensitivity\\
\hline
BR($\mu \rightarrow e \gamma$) & $<1.2 \times 10^{-11}$  &  $10^{-13}$ (MEG)\cite{MEG}\\
BR($\tau \rightarrow \mu \gamma$) & $<6.8 \times 10^{-8}$ & $ 10^{-9}$ 
(Belle)\cite{Akeroyd:2004mj}\\ 
BR($\tau \rightarrow e \gamma$) & $<1.1 \times 10^{-7}$ &\\
\hline
\hline
BR($\mu \rightarrow e \bar{\nu}_e \nu_{\mu}$) &$\sim 100 \%$& \\
BR($\tau \rightarrow \mu \bar{\nu}_{\mu} \nu_{\tau}$) &$17.36 \pm 0.05 \%$& \\
BR($\tau \rightarrow e \bar{\nu}_e \nu_{\mu}$) &$17.84 \pm 0.05 \%$& \\
\hline
\end{tabular}
\caption{Present and predicted bounds on lepton flavour violating processes, 
and measured branching ratios for $\ell_{\alpha} \rightarrow 
\ell_\beta \nu_{\alpha} \bar{\nu}_{\beta}$ decays.}
\label{table:LFVrates}
\end{center}
\end{table}

\begin{table}
\begin{center}
\begin{tabular}{|c|c|}
\hline
Present bounds (e cm)& Future sensitivity (e cm)\\
\hline
$d_e<1.6 \times 10^{-27}$  & $10^{-29}$ (Yale group)\cite{Yale}\\
$d_ {\mu}<2.8 \times 10^{-19}$ &  $10^{-24}$ (Muon EDM Collaboration) \cite{Miller:2004nw}\\ 
$(-2.2<d_{\tau}<4.5) \times 10^{-17}$&\\
\hline
\end{tabular}
\caption{Present and anticipated bounds on electric dipole moments.
See \cite{Raidal:2008jk} for a discussion of future experiments. }
\label{table:EDMbounds}
\end{center}
\end{table}

We focus on these neutrino Yukawa coupling contributions to LFV and EDMs,
assuming MSUGRA with real boundary conditions at $M_X$.
Additional contributions,
arising with less restrictive boundary conditions, 
are unlikely to cancel the ones we discuss, 
so the upper bounds that will be set if, for instance, no electron EDM is 
measured by the Yale group, will equally apply.
Conversely, if an electron EDM is measured above the range that we 
predict, it will prove the existence of a source of CP violation other 
than the neutrino Yukawa phases.

We are interested in analytic estimates
for LFV rates and electric dipole moments.  For
this,  we need  the flavour-changing and
CP violating contributions to the soft masses, 
that  arise from the neutrino Yukawa. 
Following \cite{Farzan:2004qu}, we take  the 
one-loop corrections to the flavour off-diagonal doublet 
slepton masses $\widetilde{m}^2_{L\alpha \beta} \to 
\widetilde{m}^2_{L\alpha \beta} + \Delta \widetilde{m}^2_{L\alpha \beta}$ and 
to the trilinear coupling $a_0 \lambda \to a_0\lambda (1 + \Delta a_0)$ 
to be:
\bea \label{m_soft} 
\Delta \widetilde{m}^2_{L\alpha \beta} &=& 
-\frac{1}{16 \pi^2} (3 m_0^2 + a_0^2)[C^{(1)}]_{\alpha \beta} 
-  \frac{1}{16 \pi^2} (m_0^2 + a_0^2 + 2a_0B)[H]_{\alpha \beta}~,
\\
 \label{A}
\Delta (a_0)_{\alpha \beta} &=& -\frac{1}{16 \pi^2} [C^{(1)}]_{\alpha \beta} 
               - \frac{1}{16 \pi^2} [H]_{\alpha \beta} ~,
\eea
for $\alpha \neq \beta$ where the matrices $H$ and $C^{(n)}$ are given by: 
\beq \label{H}
H \equiv \lambda \lambda^{\dagger} = V_L^{\dagger} D_{\lambda}^2 V_L ~,
\eeq
\beq \label{C}
C^{(n)} \equiv \lambda  
\log^{n}\left( \frac{M M^{\dagger}}{M_X^2} \right) \lambda^{\dagger} =  
V_L^{\dagger} D_{\lambda} V_R  \log^{n}\left( \frac{M M^{\dagger}}
{M_X^2}\right)  V_R^{\dagger} D_{\lambda} V_L ~.
\eeq
$C^{(1)}$ is the leading log contribution, and
terms $ \propto H$ arise in the finite part (they could be relevant
for EDMs).
The one loop corrections to the right handed charged slepton mass matrix, 
$\widetilde{m}^2_{R\alpha \beta}$ only contain the charged lepton Yukawa 
couplings and therefore cannot generate off-diagonal entries. These
are generated at two loops and, as we will see later, they can be relevant 
for the lepton EDMs.

At one loop, sparticles generate the dipole operator
(where $e$ without subscript is the electro-magnetic
coupling constant):
\beq
e X_{\alpha \beta} \overline{e_L} ^\alpha \sigma ^{\mu \nu}
e_R ^\beta F_{\mu \nu} + h.c. 
\label{Z1}
\eeq
 which   
 leads to LFV 
decays ($\ell_\alpha \rightarrow \ell_\beta \gamma$),
and induces the flavour diagonal  anomalous 
magnetic  and electric dipole moments of charged leptons 
\cite{Raidal:2008jk}.    For $\alpha = \beta$, the anomalous
magnetic moment is $a_\alpha = 4 m_{e_\alpha} {\rm Re}\{ X_{\alpha \alpha} \}$
and the electric dipole moment is
 $2 {\rm  Im}\{ X_{\alpha \alpha}\}$.

In the mass insertion approximation the observable LFV rates are 
proportional to $|\widetilde{m}^2_{L\alpha \beta}|^2 
\propto |C^{(1)}_{\alpha \beta}|^2$ and the corresponding branching ratios
are of order  \cite{Hisano:1995cp}:
\bea \label{radiative_decay}
\frac{BR(\ell_{\alpha} \to \ell_{\beta} ~ \gamma)}{BR(\ell_\alpha \rightarrow
\ell_\beta \nu_{\alpha} \bar{\nu}_{\beta})} &  
\sim  & \frac{\alpha^3}{G_F^2} \frac{\tan^2 \beta}{m_{SUSY}^8} 
|\widetilde{m}^2_{\alpha \beta}|^2 \\ \nonumber
 & \sim & \frac{\alpha^3}{G_F^2} \frac{\tan^2 \beta}{m_{SUSY}^8} 
\frac{(3m_0^2 + a_0^2)^2}{(4 \pi)^4} |[C]_{\alpha \beta}|^2,
\eea
where $G_F$ is the Fermi constant, $\tan \beta = v_u/v_d$, and $m_{SUSY}$ is a 
generic SUSY mass, which substitutes for the mixing
angles and the function of the loop particle masses.

An estimate of $m_{SUSY}$ can be obtained from the data on
the anomalous magnetic moment of the muon,
as suggested in \cite{Hisano:2001qz}.
A $3.3$ or $3.4\sigma$ 
deviation from the Standard Model prediction is observed in the anomalous 
magnetic moment of the muon (in Table (\ref{table:AMM}) 
is given  the  experimental  value of $a_\mu$ 
and  the deviation from the SM prediction  
\cite{Hagiwara:2006jt,Hertzog:2007hz}).
We  assume it is due to  new physics that can also contribute to flavour 
violation and EDMs. 
In the MSUGRA seesaw scenario that we are considering, the main
contribution  to $a_\mu$ comes from 
1-loop diagrams with neutralino or chargino exchange and 
is given by
\cite{Moroi:1995yh}:
\beq \label{susyAMM}
\delta a^{SUSY}_{\mu} \simeq \frac{\alpha m^2_{\mu}}{8 \pi \sin^2\theta_{weak}} \frac{\tan\beta}{m^2_{SUSY}}.
\eeq
\begin{table}
\begin{center}
\begin{tabular}{|c|c|}
\hline
$a^{EXP}_{\mu}$ & $(116~592~080 \pm 63) \times 10^{-11}$ in BNK-E821\\
\hline
\hline
 & $(276 \pm 81) \times 10^{-11}$ \cite{Hagiwara:2006jt} \\ 
$\delta a_{\mu} = a_{\mu}^{EXP}-a_{\mu}^{SM}$ & $(275 \pm 84) \times 10^{-11}$
 \cite{Davier:2007ua}\\
 & $(295 \pm 88) \times 10^{-11}$ \cite{Hertzog:2007hz}\\
\hline
\end{tabular}
\caption{Experimental value and deviation from the SM predictions of the muon 
anomalous magnetic moment. The errors of $\delta a_{\mu}$ are the  
combination in quadrature of  
the experimental and theoretical ones.}
\label{table:AMM}
\end{center}
\end{table}
{ Within this approximation, the observed deviation in the muon 
anomalous magnetic moment only fixes the ratio 
$\tan\beta/m_{SUSY}^2 \sim 5~10^{-5}~ GeV^{-2}$, so our SUSY masses scale with
$\tan\beta$ as $m_{SUSY}^2 = \frac{\tan \beta}{2} ( 200 $ GeV)$^2$.}

Assuming  \cite{Hisano:2001qz}
 that the main contribution to the LFV branching ratio is given 
by analogous diagrams involving chargino and neutralino exchange, gives, 
from equations (\ref{radiative_decay}) and (\ref{susyAMM})
with  $m_0 \simeq a_0 \simeq m_{SUSY}$:
\beq \label{BR_LFV}
\frac{BR(\ell_{\alpha} \to \ell_{\beta} ~ \gamma)}{BR(\ell_\alpha \rightarrow
\ell_\beta \nu_{\alpha} \bar{\nu}_{\beta})}\sim 10^{-8} |C_{\alpha\beta}|^2 
\left( \frac{\delta a_{\mu}}{10^{-9}} \right)^2.
\eeq
Since we aim to explore  seesaw parameter space, we
set the MSUGRA parameters $m_0 \simeq a_0 \simeq m_{SUSY}$.

In our analysis, we aim for values of $|C_{\alpha\beta}|^2$ that will give 
$\mu \to e \gamma$ and either of $\tau \to \ell \gamma$ in the next round of 
experiments. 
We require only one of the $\tau$ decays, because the other
 must be small 
to suppress $\mu \to e \gamma$ (recall that
we assume  the neutrino Yukawas are 
hierarchical).

The neutrino Yukawa corrections to the soft terms can also enhance the 
predictions of the CP violating electric dipole moments.
In our 
discussion we can neglect muon and tau EDMs, because the experimental 
sensitivity on $d_{\mu}$ is currently eight orders of magnitude weaker than 
on $d_e$ and we expect $d_{\mu}/d_e \sim m_{\mu}/m_e$. 

There are two potentially important contributions to the charged lepton EDMs
induced by the neutrino Yukawa couplings. 
As discussed in \cite{Ellis:2001yza, Farzan:2004qu}, the first non-zero
contribution to the complex, flavour diagonal
EDMs arises at two-loop order. 
The matrices $\Delta a_0$ and $\Delta\widetilde{m}_L^2$ 
in Eq.(\ref{m_soft}) are the available 
building blocks to make an EDM, which turns out to 
be proportional to the commutator $[H,C]$. This is the dominant contribution at 
low $\tan\beta$. 

We follow \cite{Farzan:2004qu} 
\footnote{\cite{Farzan:2004qu} finds the same structure as 
\cite{Ellis:2001yza,masina},
but its result is smaller by one power of a large  
logarithm.}
to estimate:
\beq
\label{edm1}
d_e \sim \frac{4 \alpha}{(4 \pi)^5}\frac{m_e^2}{m_{SUSY}^2} 
{\rm Im} [H~C]_{ee}(1.9 ~ 10^{-11}~\textrm{e cm}) \sim 10^{-29}~
\left( \frac{2}{\tan \beta} \right)
{\rm Im} [H~C]~\textrm{e cm}~,
\eeq
where we have used $[H,C]/i = 2 {\rm Im} [H~C]$, and the $2/\tan\beta$ arises because we extracted $m_{SUSY}^2$ from the $\delta a_{\mu}$.

In the large $\tan\beta$ region, it has been shown \cite{masina} that 
a different contribution to the EDMs can be the dominant one. 
This new contribution arises at three loops, and it involves the 
two loop correction to the right handed charged slepton mass matrix
$\Delta \widetilde {m}_E^2$. It is proportional to the CP violating 
quantity:
\beq 
D_\alpha = {\rm Im} \left[((\Delta\widetilde {m}_E^2)^T m 
\Delta\widetilde {m}_L^2\right]_{\alpha \alpha} 
\eeq
where $m$ is the (diagonal) charged lepton mass matrix.
Despite being a higher loop order, it is typically dominant for 
$\tan \beta \gsim 10$. The two loop expression for 
$\Delta \widetilde {m}_E^2$  can be found in \cite{Farzan:2004qu}.
We approximate this contribution as:
\bea
d_e & \simeq & \frac{-e}{2} \frac{8 \alpha}{(4 \pi)^7} 
\frac{10 m_e\tan \beta}{m_{SUSY}^2 } 
\frac{{\rm Im}[\lambda_{ek}^{*} \lambda_{\alpha k} m_{\ell_\alpha}^2 
\lambda_{\alpha m}^{*} \lambda_{e m}]}{v^2\cos^2 \beta} F(M_k^2) ,
\label{edm2}
\eea
where
\beq
F(M_k^2) =  \left( \log \frac{M_X^2}{M_N^2}  \log \frac{M_X^2}{M_k^2} 
\log \frac{M_N^2}{M_k^2}  + \log^2 \frac{M_N^2}{M_k^2} 
\log \frac{M_N^2}{M_m^2}  \right) ,
\eeq
and  $M_X  = 3 \times 10^{16}$ GeV,  $M_N = M_2$.
It gives an electric dipole moment of order:
\bea
d_e   
 & \sim  &  10^{-29}  \left( \frac{\tan \beta}{50} \right)^2
\frac{{\rm Im}[\lambda_{ek}^{*} \lambda_{\alpha k} m_{\ell_\alpha}^2 
\lambda_{\alpha m}^{*} 
\lambda_{e m}] }{m_\tau^2}
~  e {\rm cm}  . \nonumber
\eea

One comment is in order. Throughout this work, we use the approximated
formulae (\ref{BR_LFV}), (\ref{edm1}) (\ref{edm2}), where we have  
set the supersymmetric parameters $m_0$ and $a_0$ at a common $m_{SUSY}$ scale.
Of course these are very rough approximations, but given that 
a detailed analysis of the MSUGRA scenario is beyond the scope of this study, 
which concentrates on the seesaw parameters, it is enough to illustrate 
our results. 

Notice that, since we normalize the LFV branching ratios to 
the muon g-2 deviation from the SM, there is no enhancement of LFV for
large $\tan\beta$. The  three loop EDM  contribution  (\ref{edm2})
 is  enhanced, because it has  
extra powers of $\tan \beta$.

\section{Flavoured thermal leptogenesis}
\label{leptogenesis}

The observed  Baryon Asymmetry of the Universe \cite{WMAP} is:
\bea 
Y_{\Delta B} &\equiv&  \frac{n_B- n_{\bar{B}}}{s} {\Big |}_{0}
= ( 8.75 \pm 0.23)  \times 10^{-11}
\label{YB}
\eea
where $n_{B 0}$, $n_{\bar B 0}$,  and $s_0$  are the number
densities of  baryons, antibaryons, and entropy, in the Universe today.
We assume this excess is 
produced via flavoured thermal leptogenesis\cite{fy,flavour,rev}, through the 
decays of the lightest singlet 
neutrino $N_1$
 and sneutrino $\tilde{N}_1$,
in the thermal plasma at $T \sim M_1$.
The population of  $N_1$
 and  $\tilde{N}_1$ is produced by inverse decays and 
scattering in the plasma. 
The decays are CP violating and controlled by the neutrino Yukawa coupling,
 thus for hierarchical right-handed (s)neutrinos the CP-asymmetry is
 given by \cite{Covi:1996wh}:
\bea \label{eps}
\epsilon_{\alpha \alpha} 
& = &  \frac{\Gamma(N_1 \to \ell_{\alpha} H,~\widetilde{\ell_{\alpha}} h) -
\Gamma(N_1 \to \overline{\ell}_{\alpha} \overline{H},
~\overline{\widetilde{\ell_{\alpha}}} \overline{h}) 
 }{\Gamma(N_1 \to \ell H,~\widetilde{\ell} h) +
\Gamma(N_1 \to \overline{\ell} \overline{H},
~\overline{\widetilde{\ell}} \overline{h}) }
\nonumber\\
& \simeq&  \frac{3 M_1}{8 \pi v_u^2 
\left [ \lambda^{\dagger} \lambda \right]_{11}} {\rm Im}
\left \{ [\lambda]_{\alpha 1}  [m_{\nu}^{\dagger} \lambda]_{\alpha 1}  \right \} ,
\eea
where $\alpha$ specifies the flavour of the (s)lepton doublet in the final state.
If the CP violating decays are out-of-equilibrium the lepton asymmetry
 produced can survive and be partially converted into  a baryon asymmetry
 through non perturbative SM sphaleron processes\cite{Kuzmin:1985mm}.

In Eq.(\ref{eps}) we have intentionally not summed over the flavour 
index $\alpha$, because  flavours can have
 a role in the evolution of the lepton asymmetry \cite{flavour}. That is,
 if  a flavour  in the thermal bath is  distinguishable, then the corresponding
 lepton asymmetry follows an independent evolution. This occurs when the
 charged lepton Yukawa interaction 
rate
$\Gamma_{\ell_{\alpha}} = 5 \times 10^{-3} T Y_{\alpha}^2$ 
is faster than the expansion rate $H$ and the singlet inverse 
decay rate $\Gamma_{ID} \sim e^{-m/T} \Gamma_N$, where $\Gamma_N $ 
is the right-handed neutrino decay rate.
Since leptogenesis takes place at $T \sim M_1$ the mass 
of the lightest right-handed (s)neutrino tells us if flavour 
effects are important.

In the MSSM, the charged lepton Yukawas  are  larger than in
the SM: 
$Y_\alpha = m_{\alpha}/( \cos  \beta \times  174  {\rm GeV}) $, so
they come into equilibrium earlier.  
At very high temperatures 
$T>\tan^2 \beta \  10^{12} $ GeV \footnote{ We approximate  $\tan \beta
\simeq 1/ \cos  \beta $ because $ \sin  \beta \sim 1$ and 
$\tan \beta$ is a more familiar parameter.}, 
the charged lepton yukawa interactions are out of 
equilibrium ($\Gamma_{\ell_{\alpha}} \ll H $) and there are no flavour effects,
so leptogenesis  can be studied in one-flavour case. However, as 
the temperature drops,  the 
$\tau$ interactions come into equilibrium.   
  In the range  $\tan^2 \beta \  10^9 \lesssim T \lesssim 
\tan^2 \beta \  10^{12}$ GeV,  we have  an intermediate 
two-flavour regime, so that the lepton asymmetry produced in the $\tau$ 
evolves separately from the lepton asymmetry created in the linear combination:
\beq \label{lo}
\hat{\ell}_o = \frac{\lambda_{\mu 1} \hat{\mu} + 
\lambda_{e 1} \hat{e}}{\sqrt{ |\lambda_{\mu 1}|^2 + |\lambda_{e 1}|^2} } ~.
\eeq
For $T \lesssim \tan^2 \beta \  10^9$ GeV, 
 also the $\mu$ Yukawa interactions come into chemical
 equilibrium and all the three flavours become distinguishable.

In all the flavour regimes the baryon to entropy ratio can be written as:
\beq \label{bau}
Y_B \simeq \frac{10}{31}   \frac{n_N + n_{\tilde{N}}}{s}
\sum_{\alpha} \epsilon_{\alpha \alpha} \eta_{\alpha} ~ \simeq
 \frac{10}{31}    \frac{315 \zeta(3)}{4 \pi^4 g_*} 
\sum_{\alpha} \epsilon_{\alpha \alpha} \eta_{\alpha}  ~.
\eeq
The numerical  prefactor 
indicates the fraction of $B-L$ asymmetry converted into a baryon asymmetry
by sphalerons \cite{Khlebnikov:1988sr} in the MSSM. The second fraction
is the equilibrium  density
of singlet neutrinos and sneutrinos, at $T \gg M_1$,
  divided by the entropy density $s$. Numerically, it is
of order $ 4 \times 10^{-3}$, similar to the non-SUSY case \footnote{ 
The addition of the  $\tilde{N}$s  is compensated
by the approximate doubling of the degrees
of freedom in the plasma : $g_* = 228.75$ for the MSSM.}. 
The $\epsilon_{\alpha \alpha}$ are the CP asymmetries in 
each flavour (so that $\alpha = \tau, o$ or $\alpha = \tau, \mu, e$ in the two- or three-flavour 
regimes respectively) and the $\eta_{\alpha}$ are the 
efficiency  factors which take into account that these 
CP asymmetries are partially erased by inverse decays and scattering processes.
We assume the efficiency  factors   have the same
functional form and numerical factors as for non-supersymmetric 
leptogenesis \cite{flavour}:
\beq \label{wash_out}
\eta_\alpha \simeq \left[
\left( \frac{m_{*}}{2 |A_{\alpha \alpha}|\tilde{m}_{\alpha\alpha}}\right)^{-1.16}
+\left(\frac{ |A_{\alpha \alpha}| \tilde{m}_{\alpha\alpha}}{2 m_{*}}\right)^{-1}\right]^{-1} ,
\eeq
where we neglect $A$-matrix \cite{barbieri} factors 
in our numerical analysis.
The rescaled $N_1$ decay rate is  defined as :
\beq \label{mtilde}
\widetilde{m} = \sum_\alpha \widetilde{m}_{\alpha \alpha}
=   \sum_\alpha \frac{|\lambda_{\alpha 1}|^2}{M_1} v_u^2 ,
\eeq
and in supersymmetry  $m^{MSSM}_* = m^{SM}_*/\sqrt{2} 
= 4 \pi v_u^2 H_1/M_1^2 \simeq 0.78 \times 10^{-3}$ eV 
\footnote{There are factors of 2 for SUSY:  defining
$\Gamma_D$ to be the total  $N$ decay rate, we have
$\Gamma_D^{SUSY} = 2  \Gamma_D^{SM}$.
So with the definition  of eq. (\ref{mtilde}) for
$\tilde{m}$, we have $\tilde{m} =  4 \pi v_u^2 \Gamma_D^{MSSM} / M_1^2 $
as opposed to $\tilde{m} =  8 \pi v_u^2 \Gamma_D^{SM} / M_1^2 $.
So $m_*^{SUSY}  = m^{SM}_*/\sqrt{2}$, where $m_*$ is  the value
of  $\tilde{m}$ that would give  $\Gamma_D = H_1$
at $T= M_1$,
and the      factor of $\sqrt{2}$  is because there are
approximately twice as many degrees of freedom in the plasma.}, where
$H_1$ is the Hubble expansion rate at $T = M_1$.

Combining equations Eq.(\ref{bau}), Eq.(\ref{eps}), Eq.(\ref{wash_out}) and Eq.(\ref{mtilde}),
we can write the BAU as:
\beq \label{baufin}
Y_B = - \frac{10}{31} \frac{135 M_1 }{4 \pi^5 g_{*} v_u^2} \sum_{\alpha} 
\eta_{\alpha} ~
{\rm Im} \{ \hat{\lambda}_{\alpha }  
[m_{\nu}^{\dagger}  \cdot \hat{ \lambda}]_{\alpha}  \} ,
\eeq where
$\hat{\lambda}_\alpha = [\lambda]_{\alpha 1} /
 \sqrt{[\lambda^\dagger \lambda]_{11}}$.
$Y_B$  is roughly a factor of $\sqrt{2}$ larger than in the SM,
in the limit where $\tilde{m}_{\alpha \alpha} > m_*$ for all
flavours.

Supersymmetric thermal leptogenesis suffers from the so called gravitino
problem\cite{gravitinop}: in a high temperature plasma gravitinos
are copiously produced and their late decay can jeopardize successful 
nucleosynthesis (BBN). 
This gives an upper bound on
the reheat temperature of the Universe
$T_{RH}$, which  constrains  the temperature at which 
leptogenesis can take place,  and
gives an upper bound on  the singlet neutrino 
mass  $M_1 \lsim 5 T_{RH}$ \cite{pluemi, GNRRS}.
However, there is  also  a lower bound on $M_1\gsim 10^9$ GeV
\cite{di} (for hierarchical $N$s)
 to obtain a large enough lepton asymmetry.
This can be seen from (\ref{baufin}), where
$Y_B \propto M_1$. 
It has recently been suggested \cite{Giudice:2008gu}
that this conflict can be avoided
by  generating the singlet masses
after reheating. However,  we here
 assume that  $M_1 > 10^9 $ GeV
is fixed before reheating.

There are various  ways to 
obtain $T_{RH} \sim 10^9 - 10^{10}$ GeV.   If the
gravitino is unstable, the nucleosynthesis bound  leads to   
very stringent upper bounds on the reheating temperature after inflation
\cite{gravitinorec}: 
$T_{RH} \lsim 10^{4}- 10^{5}$ GeV for $m_{3/2} \lsim 10 $ TeV,
 or $T_{RH} \lsim  10^{9}-10^{10}$ GeV for  $m_{3/2} > 10 $ TeV.
A sufficiently high reheat temperature is obtained
for very heavy gravitinos because they decay before BBN. 
Alternatively, if the gravitino is the stable LSP,
a correct  dark matter relic density can be obtained
for  $T_{RH} \sim  10^{9}-10^{10}$ GeV. 
In this scenario,  one must ensure that 
the decay of the  NLSP does not  perturb BBN.
This can  be  obtained, for instance   by choosing the NLSP with care
\cite{NLSP} or by having it decay before BBN via
$R$-parity violating interactions\cite{AI}.

We can summarise that  a reheat temperature $\gsim 10^9$ GeV
is difficult but not impossible in supersymmetry. So
for the purposes of this paper, we will allow 
$M_1 < 10^{11}$ GeV.

\section{Reconstructing leptogenesis from low energy observables}
\label{Reconstructing}

In order to search for  
a connection between the low-energy observables and 
leptogenesis, we need  a  parametrisation in which
we can input the low energy observables, and
then compute the BAU.  
Ideally we want to express the high-energy parameters 
in terms of observables \cite{Davidson:2004wi}.
Therefore, we write the seesaw parameters in terms of operators acting on the
left-handed space, potentially more accessible: so we chose $D_{\nu}$, 
$D_{\lambda}$ and $V_L$ (that appears in the combination 
$\lambda\lambda^{\dagger}$) and $U_{PMNS}$. Within this bottom-up approach, 
the CP violation is now encoded in the three, still unknown, low energy phases
of the PMNS matrix $U$, and in the three unknown phases in $V_L$.
We then reconstruct the right-handed neutrino parameters in terms of 
those inputs. 

The matrices $D_{\nu}$ and $U_{PMNS}$ can be determined in {  low-energy 
 experiments}. Through neutrino oscillation experiments  we can extract the 
two neutrino 
mass differences, the PMNS matrix mixing angles and, in the future, 
the Dirac phase \cite{Bandyopadhyay:2007kx} (if Nature is kind with us).
Furthermore, we have an upper bound on light neutrino masses that comes from 
cosmological evaluations\cite{Cirelli:2006kt}, 
 Tritium beta decay\cite{Kraus:2004zw}, 
 and  neutrinoless double beta decay\cite{KlapdorKleingrothaus:2000sn}.
Observing this last process 
could  prove the Majorana nature of neutrinos and put some constraints on the 
combination of Majorana phases.

We have seen that in MSUGRA there is an enhancement of lepton flavour 
violating processes due to the neutrino Yukawa couplings. Assuming that these
processes can be measured in the near future 
constrains the coefficients $[C]_{\alpha \beta}$, 
see Eq. (\ref{radiative_decay}), which depend on $D_{\lambda}$ and $V_L$. 
We parametrise 
the $V_L$ matrix as the product of three rotations along the three axes, 
with a phase associated to each rotation:
\begin{equation} \label{V_L}
   V_L^{\dagger} =\left(
    \begin{array}{ccc}
        c_{13}^L c_{12}^L
        &  c_{13}^L s_{12}^L ~ e^{-i \rho}
        & s_{13}^L~ e^{-i \sigma}  \\
        -c_{23}^L s_{12}^L ~e^{i \rho} - s_{23}^L ~e^{-i \omega} s_{13}^L c_{12}^L~e^{i \sigma}  
        &  c_{23}^L c_{12}^L - s_{23}^L ~e^{-i \omega} s_{13}^L s_{12}^L~ e^{-i \rho} ~ e^{i \sigma}
        &   c_{13}^L s_{23}^L ~e^{-i \omega}\\
        s_{23}^L ~e^{i \omega} s_{12}^L ~e^{i \rho} - s_{13}^L c_{23}^L c_{12}^L~e^{i \sigma}
        &  -s_{23}^L ~e^{i \omega}  c_{12}^L - s_{13}^L s_{12}^L c_{23}^L ~ e^{-i \rho} ~ e^{i \sigma}
       &  c_{23}^L c_{13}^L
    \end{array} \right) ~,
\end{equation}
>From the bottom-up parameters defined above and using the equation (\ref{Dnu}),
we are now able to reconstruct the right handed neutrino mass matrix and the 
$V_R$ matrix appearing in the baryon asymmetry:
\beq \label{M-reconstr}
M^{-1} = V_R D_M^{-1} V_R^T = 
D_{\lambda}^{-1} V_L U  D_{\nu} U^T V_L^T  D_{\lambda}^{-1} v_u^{-2}~.
\eeq

In leptogenesis {\it without} flavour effects, the BAU   
is controlled only by the phases of $V_R$, 
which also contribute to the $U_{PMNS}$ in the parametrisation we use.
However, as demonstrated in the $R$ matrix parametrisation 
\cite{Casas:2001sr}, it is always 
possible to choose $V_L$ such that the lepton asymmetry $\epsilon$ has any 
value for any value of PMNS phases \cite{Branco:2001pq}.
So for $Y_B$ in its observed range,
the PMNS phases can be anything, and if we measure values of the
PMNS phases, $Y_B$ can still vanish.  
In flavoured leptogenesis, the BAU can be written as a function
of PMNS phases and unmeasurables,
but it was shown in \cite{Davidson:2007va} 
 that for the Standard Model seesaw,
$Y_B$ is insensitive to the PMNS phases.  Relations between
low energy CP violation and leptogenesis can be obtained
by imposing restrictions on
the high-scale theory, for instance that
there are no right-handed phases  \cite{Pascoli:2006ie}.

In the  case of MSUGRA,  we assume that 
we will have two more measurable quantities in the near future, 
$\mu \rightarrow e \gamma$ and 
either of $\tau \rightarrow \ell \gamma$.
Naively, we do not expect LFV rates  to add more 
information on the CP violating phases, because the rates can be 
used to  fix two (real) parameters in 
$D_\lambda$ and $V_L$.  The question is whether the 
remaining   phases and real parameters, 
can always be arranged to generate a large enough BAU.
We find the answer to be yes.
For instance, in the limit of taking only the largest neutrino 
Yukawa coupling in $D_\lambda$, the matrices $C^{(n)}$ become 
proportional to $H$, and using the parametrisation of 
the $V_L$ matrix given in Eq.(\ref{V_L}) one can 
easily see that the $CP$ violating phases of the $V_L$ matrix 
disappear from the LFV branching ratios.

Besides the LFV processes, the neutrino Yukawa couplings 
can also contribute to
 the CP violating electric dipole moments. These
contributions are expected to be below the sensitivity
of  current experiments \cite{JMR,YF}.
See  \cite{YF} for a discussion of 
the impact of EDMs on seesaw reconstruction.
In our framework with hierarchical Yukawas we expect some 
suppression on this contributions to the EDMs. 
As we have seen in Section \ref{MSUGRA}, for low $\tan \beta$ the main 
contribution is proportional to the commutator of the matrices $C^{(1)}$ 
and $H$, see eq.  (\ref{edm1}). Thus  in the limit of 
taking only the largest Yukawa, which implies $C^{(1)} \propto H$, 
the commutator is equal to zero. 
Regarding the large $\tan\beta$ regime, although the contribution to the 
EDMs has a different dependence, given in eq. (\ref{edm2}), it can be shown 
that it also vanishes in this limit.
This means that a non-zero contribution
will be suppressed by mixing angles and a smaller  eigenvalue
of $H$.

\section{Analytic Estimates }
 \label{Analytic}

If a parametrisation existed, in which
one could input the light neutrino mass matrix, the neutrino
Yukawa couplings that control lepton flavour violation, 
{\it and} the baryon asymmetry, then it would be clear
that  the BAU, and other observables,
are all insensitive to each other.  In this section,
we argue that at the minimum values of $M_1$
where leptogenesis works, such a parametrisation
``approximately'' exists.

We analytically construct a point
in parameter space that  satisfies our criteria
(large enough BAU, LFV observable soon), and where
the baryon asymmetry is insensitive  to the
PMNS phases.  To find the point, we parametrise the seesaw with
the parameters of the effective Lagrangian
relevant to $N_1$ decay.
Since the observed light neutrino mass matrix
is not an input in this parametrisation, 
one must  check 
 that the correct low energy observables
are obtained. This should occur,
in the  region of parameter space considered\footnote{ 
This area of parameter space was
also found in  \cite{DIP} using a left-handed parametrisation
inputting $W = V_L U$ instead of $V_L$. See also \cite{Vives:2005ra}.},
because  the contribution of $N_1$  to the light
neutrino mass matrix can be neglected.
We construct the point for the normal hierarchy
and small $\tan \beta$; similar constructions are possible
for the other cases.

The effective Lagrangian   for $N_1$ and $\tilde{N}_1$, at
scale $M_1 \lsim \Lambda \ll M_2$,  arises from the
superpotential:
\beq
W_{N_1} =
{ \lambda}_{\alpha 1}  L_L^\alpha  H_u N_1^c + \frac{M_1}{2} N_1^c N_1^c
+ \kappa_{\alpha \beta} ({L_L}^\alpha  H_u) ({L_L}^\beta  H_u) 
\label{W1}
\eeq 
where  $\kappa_{\alpha \beta}$  is  obtained by integrating out $N_2$ and
$N_3$.
 It is   known \cite{JosseMichaux:2007zj} that the smallest $M_1$ 
for which   leptogenesis (with hierarchical
$N_i$)   works, occurs at  $m_* \lsim \tilde{m} \lsim m_{sol}$.
So we assume that
\beq
\frac{\lambda_{\alpha 1} \lambda_{\beta 1}}{M_1} v_u^2 \ll m_{\alpha \beta} ,
\eeq
implying that 
 $N_1$ makes negligible contribution
to  light neutrino observables. 
We are therefore free to tune the $\lambda_{\alpha 1}$s
to maximise the baryon asymmetry. 

To obtain a baryon asymmetry 
$ Y_B
\simeq  10^{-3} \sum_\alpha 
\epsilon_{\alpha \alpha} \eta_\alpha
\simeq 8 \times 10^{-11} $, we require: 
\beq
 \sum_\alpha 
\epsilon_{\alpha \alpha} \eta_\alpha
 \simeq
8 \times 10^{-8} ~.
\label{epseta}
\eeq
For $\tan \beta \simeq 2$,   it is unclear whether
the $\ell_\mu$ is  distinct for leptogenesis purposes.
For simplicity we assume not, and use two
flavours $o$ and $\tau$.  The efficiency factors $\eta_\alpha$  
are  maximised to $\eta_\alpha \simeq 1/4$ 
 for   $\tilde{m}_{\alpha \alpha} = 
|\lambda_{\alpha 1}|^2  v_u^2/M_1  \simeq \sqrt{2} m_*$. 
Since $\tilde{m} \simeq 3 m_*$, this is barely
in the strong washout regime, and (\ref{wash_out}) should
be an acceptable approximation.

We would therefore like to find a point in
parameter space, such that $M_1 \sim 10^9$ GeV, 
$\epsilon_{oo} \simeq \epsilon_{\tau \tau} 
 \simeq
1.6 \times 10^{-7}$.
Defining $\hat{\lambda}_\alpha =
\lambda_{\alpha 1}/\sqrt{\sum_\alpha |\lambda_{\alpha 1}|^2}$,  
equation (\ref{eps})
implies that we need, for $\alpha  = o$ and
$\alpha = \tau$:
\beq 
{\rm Im} {\Big \{} \hat{\lambda}_{\alpha 1} 
\frac{[m^\dagger \hat{\lambda}]_{\alpha 1}}{m_3} {\Big \} } \gsim 
\frac{10^9 {\rm GeV}}{M_1} ~.
\label{Anbd}
\eeq
This means that  $\hat{\lambda}_{ 1}$ needs  a component
along $\hat{u}_3$
(the eigenvector of $m_3$), and,  since it should also
generate  $m_1$, it needs a component along  $\hat{u}_1$.
It can  always  be written as:
\beq
\vec{\lambda}_{ 1} = \lambda_{11} \hat{u}_1  +  \lambda_{21} \hat{u}_2 + \lambda_{31} \hat{u}_3  ~~~,
\label{fin5.5}
\eeq
where $\{ 1,2,3 \}$ indices indicate the light neutrino
mass basis.
 In the following  we take $ \lambda_{21} = 0$, 
$ \lambda_{31} = |\lambda_{31}| e^{i \zeta}, |\lambda_{31}|  \gg |\lambda_{11}|$. 
With equation  (\ref{m_nu_eff}),  
\bea 
{\rm Im} {\Big \{} \hat{\lambda}_{\alpha 1} 
\frac{[m^\dagger \hat{\lambda}]_{\alpha 1}}{m_3 } {\Big \} }& =
&
\frac{1}{|\lambda_{11}|^2 + |\lambda_{31}|^2 } {\rm Im} {\Big \{}
( \lambda_{11}\lambda_{31} U_{\alpha 1} + \lambda_{31}^2 U_{\alpha 3}) U_{\alpha 3}^*   {\Big \} }
\nonumber \\
&\rightarrow&
 \frac{1}{|\lambda_{11}|^2 + |\lambda_{31}|^2 }
  {\rm Im} {\Big \{}  \frac{\lambda_{31}^2}{2}    {\Big \} } 
 \nonumber%
\eea
(no sum on $\alpha$). 
In the last formula, 
we drop  the  terms  $\propto \lambda_{11}$, which may
contain
asymmetries that  cancel in the sum $\epsilon_{oo} + \epsilon_{\tau \tau}$.
These are not useful to us,
because we aim for $\eta_o \simeq \eta_\tau \simeq 1/4$. For 
{ Im} $\{ \lambda_{31}^2 \} /(|\lambda_{31}|^2 + |\lambda_{11}|^2) \gsim 1/2$, 
Eq.(\ref{Anbd}) implies that  a large enough BAU could
be produced for $M_1 \sim  3 \times 10^9$ GeV. 

We now check that we  obtain the observed light
neutrino mass matrix, even with  $\zeta$,
the phase of $\lambda_{31}$, of order $\pi/4$. 
The light neutrino mass matrix is:
\beq
[m]_{\alpha \beta} = \frac{\lambda_{\alpha 1} \lambda_{\beta 1}}{M_1} v_u^2 +
\kappa_{\alpha \beta} v_u^2
=  v_u^2 { \Big [ } \frac{\lambda_{11}^2}{M_1} \hat{u}_1\hat{u}_1^T
+ \kappa_2  \hat{u}_2\hat{u}_2^T
+
 (\frac{\lambda_{31}^2}{M_1}  + \kappa_3 ) \hat{u}_3\hat{u}_3^T {\Big ]} _{\alpha \beta}
\eeq
where $\kappa_2$ and $\kappa_3$ are the eigenvalues
of $\kappa$. 
By convention there is no phase on
$m_3$,    so in the 2 right-handed neutrino 
(2RHN) model that generates $\kappa$, we should
put a phase on the larger eigenvalue  $\kappa_3$.
Since $\lambda_{31}^2v_u^2/M_1 \simeq  e^{i2 \zeta} \times 10^{-3}$ eV,
the phase  on $\kappa_3$ is  very small and we neglect
it in the following discussion of lepton flavour violation.. 
It is well known \cite{2RHN}  that  the seesaw
mechanism with 2 right-handed neutrinos can reproduce
the observed light neutrino mass matrix, with $m_1 = 0$.
In our case, we assume that $N_2$ and $N_3$  give
the observed $m_{2}$,  and $m_3$  up to  (negligeable)
corrections  due to $N_1$   of order $10^{-3}$ eV.  
$m_1$ arises due to $N_1$.

In the 2RHN model,
there is less freedom to tune the LFV branching
ratios \cite{Ibarra:2003up} than in the seesaw
with three $N_i$.  So as a last step, 
 we  check  that  we can  obtain LFV branching
ratios just below  the current sensitivity.
 The 2RHN model  can  be conveniently
 parametrised with $\hat{D}_\kappa,$ the $ 3 \times 2$  $ \hat{U}_{PMNS}$ matrix,
the $2 \times 2$  unitary matrix $ \hat{W} = \hat{V}_L \hat{U}$, 
and the eigenvalues 
 $\Lambda_2$ and  $\Lambda_3$ of 
$\hat{\Lambda}$ (matrices in the 2RHN
subspace are denoted by hats). 
$\hat{\Lambda}$ is a   $2 \times 2$ sub-matrix of  $\lambda$,
obtained by   expressing the $3 \times 3$ Yukawa matrix
in the eigenbases of the heavy and light neutrinos,
and dropping the first row and column, corresponding
to $\nu_1$ and $N_1$. It is straightforward  to
verify that
 $[\hat{V}_L]_{3 e} \sim 10^{-3} $ can be obtained by taking
 $\tan \hat{\theta}_W \simeq  s_{13}/(c_{13} s_{12})$,
where $ \hat{\theta}_W$ is  the rotation angle in
$\hat{W}$ and $\theta_{ij}$ are from $U_{PMNS}$.
Choosing  $\Lambda_2$, 
the smaller eigenvalue of $\Lambda$,
  to be $\sim .06$,  ensures
that $BR(\mu \to e \gamma)$ is small enough.   We can
simultaneously take $\Lambda_3 \sim 1$ and obtain
 $[V_L]_{3 \tau} \sim [V_L]_{3 \mu} \sim 1 $, which
allows  $BR(\tau \to \mu \gamma) \sim 10^{-8}$.
The resulting masses of $N_2, N_3$ are $\sim 10^{12}, 10^{15}$ GeV.

Our MCMC has some difficulties in finding the analytic points. 
We imagine this to be because they
are ``fine-tuned''  in the parametrisation used by the MCMC.
The amount of tuning required in the angles of $V_L$, to
obtain the desired $\{ \lambda_{j1}  \}$, can be
estimated by taking logarithmic derivatives. 
In Appendix \ref{Appft},
we find a  fine-tuning  of order:
\beq
\frac{\tilde{m}^2}{m_3^2\theta_{13}} \sim 
.01 
\eeq
where $\theta_{ij}$ are the $U_{PMNS}$ phases, and
we  optimistically assumed $\theta_{13} \simeq .1$.
These points at $M_1 \lsim 10^{10}$ GeV with
$\tilde{m} \gsim 10^{-3}$ eV, were also not found
in the analysis of \cite{Ellis:2002xg}.

\section{MCMC}
\label{MCMC}

In this section we describe our numerical analysis. In order to verify if the baryon asymmetry of the universe is sensitive to the 
low energy PMNS phases, we perform a scan over the neutrino sector parameters 
aiming for those points compatible with the measured baryon asymmetry and the 
bound on the reheating temperature, that have large enough LFV branching 
ratios to be seen in the next experiments.

Using the bottom-up parametrisation of the seesaw defined by the 
$V_L$, $D_\lambda$, $D_{\nu}$ and $U$ matrices, 
our parameter space consists of the 14
variables displayed in Table \ref{table:param}. 
We take as an experimental input 
the best fit values of the light neutrino mass differences and of the solar 
and atmospheric mixing angles, Table \ref{table:neutrinofit}.
With respect to the SUSY parameters, we choose two different regimes for $\tan\beta$, equal to $2$ or $50$, while the $m_{SUSY}$ scale is deduced from the data on the anomalous magnetic moment, see section \ref{MSUGRA}.

\begin{table} 
\begin{center}
\begin{tabular}{|c|}
\hline
Light neutrino best fit values\\
\hline
$\Delta m_{sol}^2 = (7.60 \pm 0.20) \times 10^{-5} ~\textrm{eV}^2$ \\
$|\Delta m_{atm}^2| = (2.40 \pm 0.15) \times 10^{-3} ~\textrm{eV}^2 \nonumber$ \\  
$\sin\theta_{sol}^2 = 0.320 \pm 0.023$ \\
$\sin\theta_{atm}^2 = 0.500 \pm 0.063$ \\
\hline
\end{tabular}
\caption{The best fit values of the light neutrino parameters and their 
$1\sigma$ errors \cite{Maltoni:2004ei}.}
\label{table:neutrinofit}
\end{center}
\end{table}
Due to the large number of parameters it would prohibitive to consider a usual grid scan.
Thus, we choose to explore our parameter space by a Markov Chain Monte Carlo that behaves much more efficiently,
and has been already successfully employed in other analyses \cite{MCMC}.

\subsection{Bayesian inference}

Given a model with free parameters $X = \{x_1, \dots, x_n\}$ and a set of derived parameters $\xi(X)$, for an experimental data set $d$, the central quantity to be estimated is the \emph{posterior distribution} $P(X|d)$, which defines the probability associated to a specific model, given the data set $d$. Following the Bayes theorem, it can be written as:
\beq \label{eq:Bayes}
P(X|d) = \frac{\mathcal{L}(d|\xi(X)) \pi(X)}{P(d)}~,
\eeq
where $\mathcal{L}(d|X)$ is the well known likelihood, that is the probability of reproducing the data set $d$ from a given model $X$, $\pi(X)$ is the prior density function, which encodes our knowledge about the model, and $P(d) = \int \mathcal{L}(d|\xi(X)) \pi(X)dX$ is an overall normalization neglected in the following. In the case of flat priors:
\beq \label{eq:flatprior}
\pi(X) = \left\{\begin{array}{ll}
\frac{1}{X_{max}-X_{min}} & ~~\textrm{if}~~ X \in [X_{min},X_{max}] \\
0 & ~~\textrm{otherwise}
\end{array}
 \right.
\eeq
 the posterior distribution reduces to the likelihood distribution in the allowed parameter space.

The main feature of the Markov chains is that they are able to reproduce a specific target distribution we are interested in, in our case the posterior distribution, through a fast random walk over the parameter space.
The Markov chain is an ordered sequence of points $X_i$ with a \emph{transition probability} $W(X_{i+1}|X_i)$ from the $i-th$ point to the next one.
The first point $X_0$ is randomly chosen with prior probability $\pi(X)$. Then a new point is proposed by a \emph{proposal distribution} $Q(X_{i+1}|X_i)$ and accepted with probability $\mathcal{A}(X_{i+1}|X_i)$. The transition probability assigned to each point is then given by $W(X_{i+1}|X_i) = Q(X_{i+1}|X_i)\mathcal{A}(X_{i+1}|X_i)$.
Given a \emph{target distribution} $P(X)$, if the following \emph{detailed balance condition}:
\beq \label{eq:detailed_balance}
W(X_{k}|X_j)P(X_j) = W(X_j|X_k)P(X_k)
\eeq
is satisfied for any $j,k$, then the points $X_i$ are distributed according to the target distribution. For a more detailed discussion see \cite{MacKay,Gilks}.  

\subsection{The Metropolis-Hastings algorithm}

In order to generate the MCMC with a final posterior distribution (\ref{eq:Bayes}), we use the Metropolis-Hastings algorithm.
In the following, we briefly recall how the algorithm behaves, but the discussion is done in terms of the likelihood, instead of the posterior distribution, since we assume flat priors on our parameter space, see eq. (\ref{eq:flatprior}).

Let $X$ be the parameter set we want to scan, and $\mathcal{L}(X)$ 
our likelihood function, the target distribution.
>From a given point in the chain $X_i$ with likelihood $\mathcal{L}(X_i)$, a 
new point $X_{new}$
with likelihood $\mathcal{L}(X_{new})$ is randomly selected by a gaussian proposal distribution $Q(X_{new},X_i)$ centered in $X_i$ and having width $\epsilon$.
This last quantity $\epsilon$ controls the \emph{step size} of the random walk. The new point is surely
added to the chain if it has a bigger likelihood, otherwise the chain adds the new point with probability 
$\mathcal{L}(X_{new})/\mathcal{L}(X_i)$ . So the value of the next point 
$X_{i+1}$ in the chain is 
determined by:
\beq
X_{i+1} = \left \{
\begin{array}{cc}
X_{new} & {\rm with ~ probability} ~~ \textrm{min}[\mathcal{A}(X_{new}, X_i),1] \\
X_{i} & {\rm with ~ probability} ~~ 1 - \textrm{min}[\mathcal{A}(X_{new}, X_i),1]\\
\end{array} \right. ~,
\eeq
where $\mathcal{A}(X_{new}, X_i)$ is the acceptance probability:
\beq
\mathcal{A}(X_{new}, X_i) = \frac{\mathcal{L}(X_{new})}{\mathcal{L}(X_i)}   ~.
\eeq

 Given this acceptance distribution and using the symmetry of our proposal distribution  $Q(X_l,X_i)$ under the exchange $l \leftrightarrow i$, it is straightforward to see that the detailed balance condition \ref{eq:detailed_balance} is satisfied for the likelihood $\mathcal{L}(X)$ as target distribution. This implies that when the chain has reached the equilibrium, after a sufficiently long run, our sample is independent of the initial point and distributed according to $\mathcal{L}(X)$.

In order to arrive at the equilibrium in a reasonable amount of time,
the step scale $\epsilon$ of our random walk must be accurately chosen.
Indeed, if we define the acceptance rate as the number of points accepted over the number of points proposed, a too big step $\epsilon$ implies a too low acceptance rate, so that our Markov Chain never advances, while a too small $\epsilon$ and, so, a very large acceptance ratio, implies that our chain needs a very large time to scan all the space. It has been suggested that $\epsilon$ must be chosen according to an optimal acceptance rate between $20 \%$ and $50 \%$.
However, in order to ensure the detailed balance condition, $\epsilon$ cannot change during the run of the chain, thus, it is set by our program in a burn-in period.

A valid statistical inference from the numerical sample relies on 
the assumption that the points are distributed according to the target 
distribution. The first points of the chain are arbitrarily chosen and the 
chain needs a burn-in period to converge to the target distribution. 
The length of the burn-in strongly depends on the intrinsic properties of the 
chain and cannot be set \emph{a priori}. It changes according to the complexity
 of the model, to the target distribution, and the efficiency of the proposal distribution employed. Once the chain has reached the equilibrium the first 
burn-in points must be discarded to ensure the independence of the chain from the initial conditions.
Nevertheless, as we will see in section \ref{convergence}, even following the procedure above, it can be a delicate issue to determine if a chain has really converged.

\begin{table}[t!]
\begin{center}
\begin{tabular}{|l|l|l|}
  \hline  {\bf Free parameters}   & \multicolumn{2}{|l|}{{\bf Allowed range}  $[X_{min},X_{max}]$}
\\ \hline  & $ \lambda_2/\lambda_1 \simeq \lambda_3/\lambda_2 \simeq 30$ & $ \lambda_2/\lambda_1 \simeq 100$, $\lambda_3/\lambda_2 \simeq 50$
\\ \hline  $\log_{10}\lambda_3$ & $[-0.3,0.3]$ &  $[-0.5, 0.5]$
\\ \hline  $\log_{10}\lambda_2$ & $[-1.77,-1.17]$ & $[-2.2,-1.2]$
\\ \hline  $\log_{10}\lambda_1$ & $[-3.25,-2.65]$ & $[-4.2,-3.2]$
\\ \hline \hline $\log_{10}(m_1/\textrm{eV})$ & \multicolumn{2}{|l|}{$[-6,-3]$}
\\ \hline  $\log_{10}\theta_{ij}^{V_L}$ & \multicolumn{2}{|l|}{$[-4,\log_{10}\pi]$}
\\ \hline  $\rho, \omega, \sigma$ &  \multicolumn{2}{|l|}{$[0,\pi]$}
\\ \hline  $\theta_{13}$ & \multicolumn{2}{|l|}{$[0.,0.2]$}
\\ \hline  $\delta$ & \multicolumn{2}{|l|}{$[0,\pi]$}
\\ \hline  $\alpha, \beta$ & \multicolumn{2}{|l|}{$[0,\pi/2]$}
\\ \hline  
\end{tabular}
\caption{Allowed parameter space, so that the uniform prior on each parameter is defined as in eq.(\ref{eq:flatprior}).}
\label{table:param}
\end{center}
\end{table}

\subsection{The seesaw sample}
\label{sample}

In our work the free variables $X$ are given by the $14$ seesaw parameters, with uniform priors, Eq.~\ref{eq:flatprior}, on the allowed range of parameter space (see Table \ref{table:param}). The choice of a logarithmic scale on some unknown parameters allows us to scan with the same probability different orders of magnitude.
We analyze models with two different hierarchies in the neutrino Yukawas, so that, for a $\lambda_3 \sim 1$ we impose
$ \lambda_2/\lambda_1 \sim \lambda_3/\lambda_2 \sim 30$ or  $ \lambda_2/\lambda_1 \sim 100$ and $\lambda_3/\lambda_2 \sim 50$.
The lightest neutrino mass is allowed to vary between three orders of 
magnitude $10^{-6} < m_1 < 10^{-3} ~eV$ and the $\theta_{13}$ mixing angle 
within its $3\sigma$ range, $0 < \theta_{13} < 0.2 ~ \textrm{rad}$.
The $V_L$ mixing angles can vary over 4 orders of magnitude, with maximum value $\pi$.
All the CP violating phases, those of the $V_L$ matrix indicated by $\rho$, 
$\omega$ and $\sigma$ and the Dirac and Majorana phases $\delta$, $\alpha$ and 
$\beta$, are allowed to vary on all their definition range: $[0, \pi/2]$ for 
the Majorana phases and $[0, \pi]$ for the others (this avoids degeneracies).

The idea is, now, to generate a sample of points in our parameter space that 
provide enough BAU, give LFV rates big enough to be seen in the next 
generation of experiments, and also have an $M_1$ light enough to avoid 
the gravitino problem. We then define our set of derived parameters $\xi(X)$ 
as in Table \ref{table:derivedparam} and we associate to them a multivariate 
gaussian likelihood with uncorrelated errors:
 \beq \label{likelihood}
\mathcal{L}(\xi_{exp}|\xi) = \frac{1}{(2 \pi)^{1/2} \mathcal{R}^{m/2}}~
\textrm{exp}\{-\frac{1}{2}(\xi-\xi_{exp})^t \mathcal{R}^{-1} (\xi-\xi_{exp}) 
\} ~.
\eeq
Where $m = 4$ is the dimension of the derived parameter set. 
The centre values $\xi_{exp}$ are the best fit values and $\mathcal{R}$ is an 
$m \times m$ error matrix, in this case diagonal, since we assume no 
correlation between the errors. As we can see in Table \ref{table:derivedparam}, the BAU is set to its experimental value, while the LFV rates are set to be 
one order of magnitude below the present bounds, and the expected value of 
lightest heavy neutrino mass $M_1 \sim 10^{9}$ GeV is set to escape the 
gravitino problem. The branching ratio of LFV $\tau$ decays is given in terms 
of the combination $BR(\tau\to e\gamma) + BR(\tau\to \mu\gamma) 
\equiv BR_{\tau \alpha}$, since one of them is suppressed to respect the 
stringent bound from $BR(\mu\to e \gamma)$ (we assume hierarchical yukawas).

For each point $X_i$ of the chain, the lepton flavour violating branching 
ratios are estimated with equation 
Eq.(\ref{BR_LFV}), while $Y_B$ is computed after the reconstruction of the 
right neutrino mass, see Eq.(\ref{M-reconstr}), using Eq.(\ref{bau}) 
 in the flavour regime is in act at the temperatures we consider. 
We recall that the temperature at which leptogenesis takes place is of the same order of the reconstructed right-handed neutrino mass.
Depending on the value of $\tan\beta$, the range of temperatures at which the 
flavour regimes have a role changes. 
As we already mentioned in Section \ref{leptogenesis}: for small 
$\tan \beta$, in the temperature range $10^9~\textrm{GeV} < T < 10^{12}~\textrm{GeV}$ the 
$\tau$ flavour is in equilibrium and the two flavour regime is in order; 
while for $T < 10^9~GeV$ $\mu$ are also in equilibrium and
the three flavours are distinguishable. Since we aim for values of $M_1 \sim 10^9~\textrm{GeV}$ if we consider a small value of 
$\tan \beta$ our program takes into account that the BAU can be produced in both two or three flavour regimes.
For very large $\tan \beta$, instead, already for $T < 10^{12}~\textrm{GeV}$ $\tau$ and $\mu$ are 
in equilibrium, thus the three flavour regime always takes place.

In the case of steeper yukawa hierarchy, in agreement with our analytical 
estimate, we enlarge our set of derived parameters and maximise the 
rescaled $N_1$ decay rate to $\tilde{m} \sim 10^{-3}$ eV 
and the heaviest right-handed neutrino masses to  $M_2 \sim 10^{12}$ GeV and 
$M_3 \sim 3~10^{14}$ GeV.

All the points that do not respect the present bounds on LFV, do not have 
large enough baryon asymmetry or have $M_1>10^{11}$ GeV, have a null 
likelihood.  We assume that the largest uncertainty on the baryon asymmetry
comes from our calculation, so we allow $Y_B$ to be as small as 
$4~10^{-11}$. Those points having one of the RH neutrino masses above the 
$M_{GUT} \sim 3~10^{16}$ GeV scale have a null likelihood too, since in that 
case the equations we use for the evaluation of LFV processes do not apply.

\begin{table}[t!]
\begin{center}
\begin{tabular}{|l|l|}
\hline  Derived parameters $\xi(X)$  &   $\xi_{exp} \pm \sigma$ 
\\ \hline  
 $Y_B$ & $(8.75 \pm 0.23)~10^{-11}$
\\ \hline $\log_{10}BR(\mu->e \gamma)$ & $-13 \pm 0.1$
\\ \hline $\log_{10}BR(\tau-> l\gamma) $ & $-9 \pm 0.1$
\\ \hline  $\log_{10}(M_1/GeV)$ & $-9 \pm 0.1$
\\ \hline  
\end{tabular}
\caption{Best values and errors for the derived parameters $\xi(X)$ we want to 
maximize.}
\label{table:derivedparam}
\end{center}
\end{table}

\subsection{Convergence}
\label{convergence}

Convergence of the chain ensures the sample is distributed according to the target distribution and thus allows to be confident of its statistical information.
The question we want to answer in this paper, however, does not require a statistical 
interpretation of the sample. Here we only aim to show that, for any value of the 
low energy phases, the unmeasurable high energy parameters
can be rearranged to obtain the right baryon asymmetry. 
Therefore a careful diagnostic of the convergence is not a priority. 
Nevertheless, we briefly discuss it in this section since it is an important 
issue that can help the reader to have a better overview on our results.
Our sample, indeed, has some typical features that can make difficult to check if the chain has reached the target distribution.

As a rudimentary attempt, in our analysis we use the simplest and straightforward approach.
We run different chains starting from different values and compare the behaviour of the free parameters, once the chains have converged they should move around the same limiting values. However, this method can be inadequate in case of \emph{poor mixing}, i.e. when the chains are trapped in a region of low probability relative to the maximum of the target distribution.
This happens in models with strongly correlated variables,
 when the proposal distribution does not efficiently
 escape this region.
Therefore, it can be an issue for  our numerical analysis, when, 
as mentioned in section \ref{Analytic}, 
we look for a  fine-tuned region  with  a large 
baryon asymmetry and  low $M_1$. 
We can understand the poor mixing situation if we imagine a landscape on the parameter space corresponding to the target distribution, with some broad hills and a tall but very thin peak at the maximum of the target distribution. In that case, the step of the chain can be optimized to efficiently scan all the space but, if its size is larger than the width of the peak, it can easily miss it.

In case of strongly correlated variables it can also happen that the
region to be scanned is mainly a plane, that is with almost null
likelihoods. This is the case of our sample, where we expect a large region with null or almost null likelihood, for all those points that do not have large enough baryon asymmetry, low $M_1$ or do not respect the bounds on LFV.  In this context,
if a gaussian-like proposal distribution, as in our sample, is employed, the choice of the starting point becomes important to allow the chain to advance. Indeed, if the initial value is surrounded by points with null likelihood (and so null acceptance rate) and its distance from the interesting region is much larger than the step of the random walk, the chain cannot move from this point, since it always finds points with null likelihood. On the other side, if the chain starts in a region which is a reasonable fit to the data, it advances. Discarding the first points of the chain can ensure independence of the chain of the initial conditions inside the interesting region however, if this region is well separated from another interesting region, the chain has almost null probability to find the second one.

In order to perform a valid statistical analysis, more sophisticated methods should be employed to decide if the chain has converged.
In literature many studies exist on convergence criterion that help to check the mixing of the sample and are based  on the similarity of the resulting 
sampling densities of input parameters from different chains. An example can be found in \cite{ALWMCMC} and \cite{MCMCWMAP}.

\subsection{Run details}
\label{run}

In this subsection we explain the details of our MCMC run.
The parameter space we scan is very large if compared to the derived variables and, in addition, we expect a strong correlation between the evaluated baryon asymmetry and the lightest right-handed neutrino mass, see eq. \ref{baufin}.
Thus, since we expect a sample with poor mixing, as discussed in section \ref{convergence}, we first look for an initial point
which is a reasonable fit to our observables. 
This procedure is done running previous shorter chains without imposing null likelihoods to the not interesting points.
Once a wide enough set of interesting starting points is found, we start 
running the chains. 

All the simulations we present are performed by running $5$ chains with $10^6$ points each. As explained
before, during the first burn-in iterations, the scale of the random walk $\epsilon$ is varied until 
the acceptance rate of points is between the optimal range $20\%$ and $50\%$. This usually
takes much less than $3~10^3$ iterations. When the optimal acceptance rate is reached, 
the scale $\epsilon$ is fixed during the rest of the run.
The chains are then added together after having discarded the first $10^5$ points, corresponding to the burn-in period, in order to give enough time to the chain to converge.
As discussed above, this procedure should eliminate the
dependence on the initial point inside the interesting region, but is only a first attempt to ensure the sample has reached equilibrium.
We run simulations for both 
normal and inverted hierarchy, in the two cases of small and large $\tan\beta$.

\section{Discussion}
\label{Discuss}

\subsection{Assumptions}
We assume  a three generation type I seesaw  with  a hierarchical
neutrino Yukawa matrix.   We require
that  this model    produces the baryon asymmetry
via flavoured  thermal leptogenesis, and induces the 
observed light neutrino mass matrix. 
This model  has a hierarchy problem, so we 
include  supersymmetry. 

We  make a number of approximations and assumptions in supersymmetrising
the seesaw. First,  we use  real and 
universal soft terms at some  high scale,
above the masses  $M_i$ of the singlet neutrinos. 
In  this
restrictive model, 
the only contributions to 
 flavour off-diagonal
elements   of  the slepton mass$^2$ matrix
 $ \equiv [\tilde{m}^2]_{\alpha \beta}$, arise 
due to Renormalisation Group running. 
Second, we use simple leading log estimates 
for the  off-diagonals $ [\tilde{m}^2]_{\alpha \beta}$.
Third, we estimate the SUSY contributions to the
dimension five dipole operator (see Eq.\ref{Z1})
using simple formulae of dimensional analysis (see
equations (\ref{BR_LFV}),(\ref{edm1}), (\ref{edm2})).  
This operator induces flavour diagonal 
electric and magnetic dipole moments, and
the flavour changing decays $\ell_\alpha \to
\ell_\beta \gamma$.  We assume 
 the  $(g-2)_\mu$ anomaly is due to supersymmetry,
and use it to ``normalise''  the dipole
operator.  This  implies that our SUSY
masses scale with $\tan \beta$:
$m_{SUSY}^2 = \frac{\tan \beta}{2} ( 200 $ GeV)$^2$.
We imagine that there is
an uncertainty  $\sim 10$ in our estimates of electric
dipole moments  and $\ell_\alpha \to \ell_\beta
 \gamma$  decays rates,
due to mixing angles and sparticle mass differences.

Our first approximation, of universal soft terms,
seems contrary to our phenomenological perspective:
the  RG-induced  
contributions to  $ [\tilde{m}^2]_{\alpha \beta}$  can be interpreted as
 lower bounds on the mass$^2$ matrix elements. 
However, 
 we   neglect other contributions, and
 require that  the  RG induced 
flavour-violating mass  terms are $\propto C^{(1)}_{e \mu}$
(see eq. (\ref{C})),
give  detectable rates for 
$\mu \to e \gamma$ and $\tau \to \ell \gamma$ 
 in  upcoming experiments. 
Realistically,   measuring $\mu \to e
\gamma$  mediated by sleptons might allow to
determine $\tilde{m}^2_{e \mu}$, but 
does not determine the seesaw model
parameters  $C^{(1)}_{e \mu}$.
This model dependence  is compatible with our
phenomenological approach,  because our result
is negative: we say that {\it even if}
we could determine   $C^{(1)}_{e \mu}$,  
the baryon asymmetry  is
insensitive to the PMNS phases. 

In our numerical analysis we sample the lightest
neutrino mass $m_1$ and the PMNS mixing angle $\theta_{13}$, 
but these two low energy parameters could be eventually measured.
In this case our simulations should be reconsidered. However, from the 
analytical estimates, we do not expect that fixing these parameters will
change our conclusions.

\subsection{Method}
\label{method}
We explore the seesaw parameter space with a Monte
Carlo Markov Chain, for two reasons.
First, an MCMC is  more efficient than a grid scan
for multi-dimensional parameter space. 
It is essentially a programme for exploring hilltops 
in the dark. Since the programme { likes to }step up and 
is reluctant to step down,  it takes most of its
steps in the most probable areas of parameter space. 

The second potential advantage of a MCMC, is that
it could make the results less dependent on   the  priors,
that is, the choice of seesaw parametrisation,
and  of the distribution of points.  
The results of parameter space scans
are often presented  as scatter plots,
and it is difficult to not
interpret  the point distribution as probability.  
However, the density of points in
the scatter plots depends not only on what the
model predicts, but also on the distribution
of input points. For this reason,
seesaw scans using different parametrisations
can distribute points  differently in scatter plots.
For example, if a model parameter such as a
Yukawa can vary between 0 and 1, 
the results will be different depending
on whether the Yukawa is ${\cal O}(1)$
(take points uniformly distributed between
0 and 1) or  can vary by orders
of magnitude (take the exponential of a variable
uniformly distributed between $-n$ and 0).
We had  hoped  that an MCMC could improve this, because a converged MCMC
distributes points in parameter space according to  a likelihood
function.  
 However,  in practise there are various difficulties.

The  prior on  the seesaw  model parameter
space  matters, because the MCMC
takes steps of  some size in each parameter: broad hilltops
are easier to find than  sharp peaks.
As discussed in   \cite{MCMCWMAP},
this can be addressed   by
describing the model with parameters 
that match closely   to physical observables. 
For this reason we  parametrise
the seesaw in terms of 
the diagonal singlet mass matrix $D_M$, the light
neutrino mass matrix $m = U D_\nu U^T$,  and the neutrino
Yukawa matrix $\lambda \lambda ^\dagger = V_L^\dagger D_\lambda^2 V_L$.
These are  related to low energy observables, because 
 $\lambda \lambda ^\dagger$ controls the  RG contributions
to the slepton mass matrix.  
We take the priors for our inputs as given in Table \ref{table:param}.
However, the baryon asymmetry  and the  mass $M_1$   
 belong to the ``right-handed''  sector, so 
 are   complicated functions of the  ``left-handed'' input parameters.
The bridge between the LH and RH sector is
the Yukawa matrix, whose hierarchies may
strongly distort  the MCMC step size.   
To obtain a large enough baryon asymmetry for
$M_1 \sim 10^9$ GeV requires careful tuning in
the ``right-handed'' space,
and our MCMC has difficulty to find these points. 
This is related to a second, practical  problem,  that 
 there are many more parameters than observables,
so the  space to explore is big, but
the peaks with enough baryon asymmetry and
small enough  $M_1$ are rare.
It is difficult
to ensure that the MCMC has found all the peaks,
as is discussed in section \ref{convergence}. 

In section \ref{Analytic}, we find  analytically
an area of parameter space that satisfies
our constraints, but where  the baryon
asymmetry is  insensitive to  PMNS phases. This area corresponds to the limit
where  $N_1$  makes a negligible contribution
to the light  neutrino mass matrix.
In this area,
the seesaw model can be conveniently  parametrised
with the interactions of the effective theory
at $M_1$, and   it is straightforward to tune the  coupling constants to  fit
the light neutrino mass matrix, LFV rates, and
the baryon asymmetry.

\subsection{Results}

The aim of our analysis was to verify 
if a preferred range of values for PMNS 
phases $\delta$, $\alpha$ and $\beta$
 can be predicted,  once low energy neutrino  oscillation data, 
a large enough BAU, and LFV processes within the sensitivity 
of future experiments are requirements of the model. 

\begin{figure}[t!]
  \begin{center}
    \includegraphics[width = 0.495\textwidth]{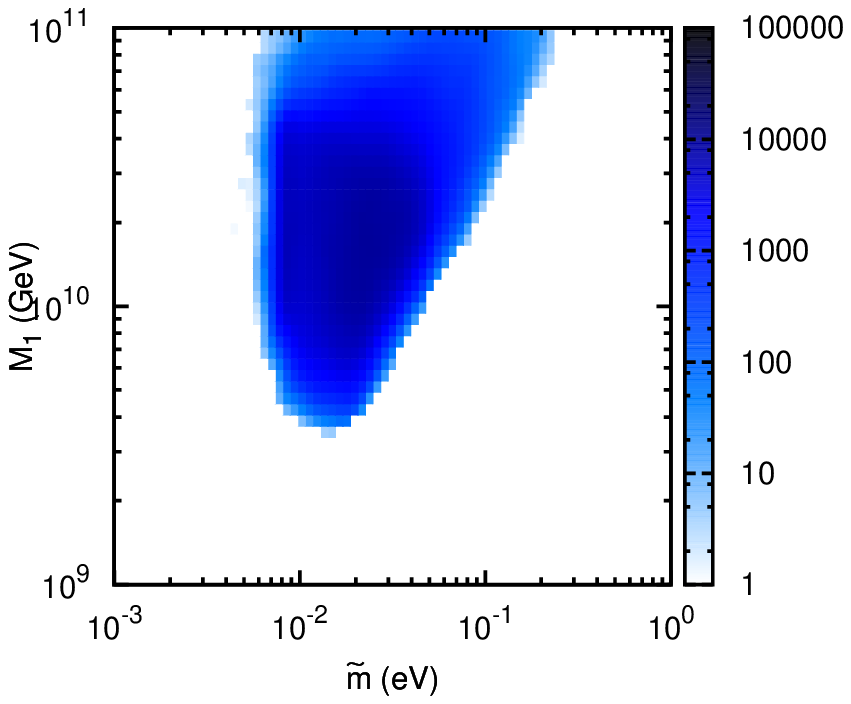}
    \includegraphics[width = 0.495\textwidth]{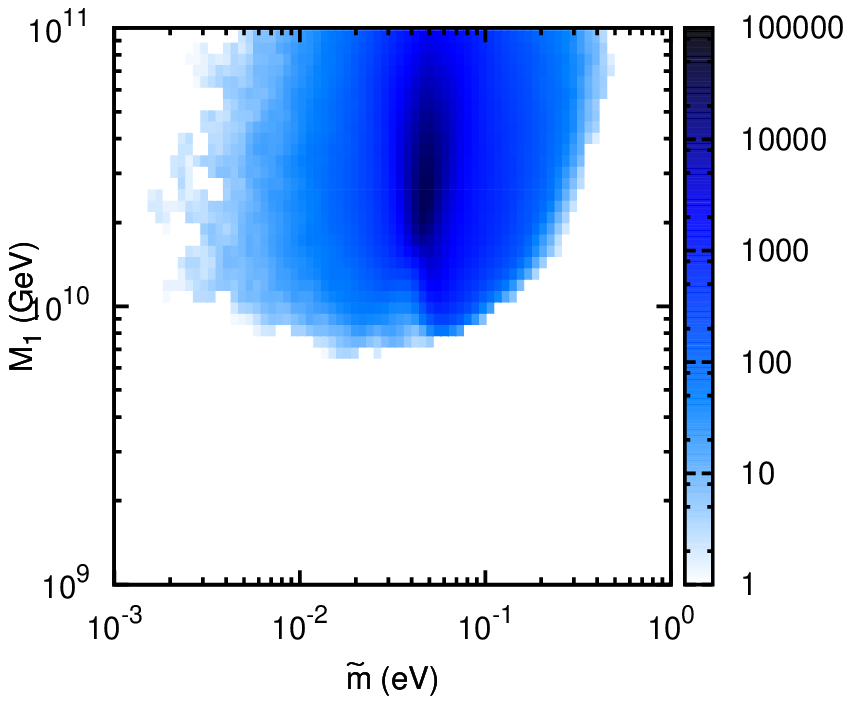}
  \caption{
Density of ``successful'' points, as a function of 
the lightest right-handed neutrino mass $M_1$ and rescaled decay rate $\tilde{m}$, assuming $\lambda_3 \sim 1$ and
$ \lambda_2/\lambda_1 \sim \lambda_3/\lambda_2 \sim 30$, for two different simulations: NH and $\tan \beta=50$ (left), and IH and $\tan \beta=2$ (right).
``Successful'' points have $Y_B > 4~10^{-11}$, 
and $BR(\mu \to e \gamma)$
and $BR(\tau \to \ell \gamma)$ an order of magnitude below
the current bounds. See section \ref{sample}. }
  \label{Fig:M1_mt}
  \end{center}
\end{figure}

In Fig. \ref{Fig:M1_mt}, we show the distribution, as a function of the singlet
 mass $M_1$ and the (rescaled) decay rate $\tilde{m}_1$,
 of the successful  points for a yukawa hierarchy $\lambda_2/\lambda_1 \sim \lambda_3/\lambda_2 \sim 30$, with $\lambda_3 \sim 1$.  
%

\begin{figure}[t!]
  \begin{center}
    \includegraphics[width = 0.495\textwidth]{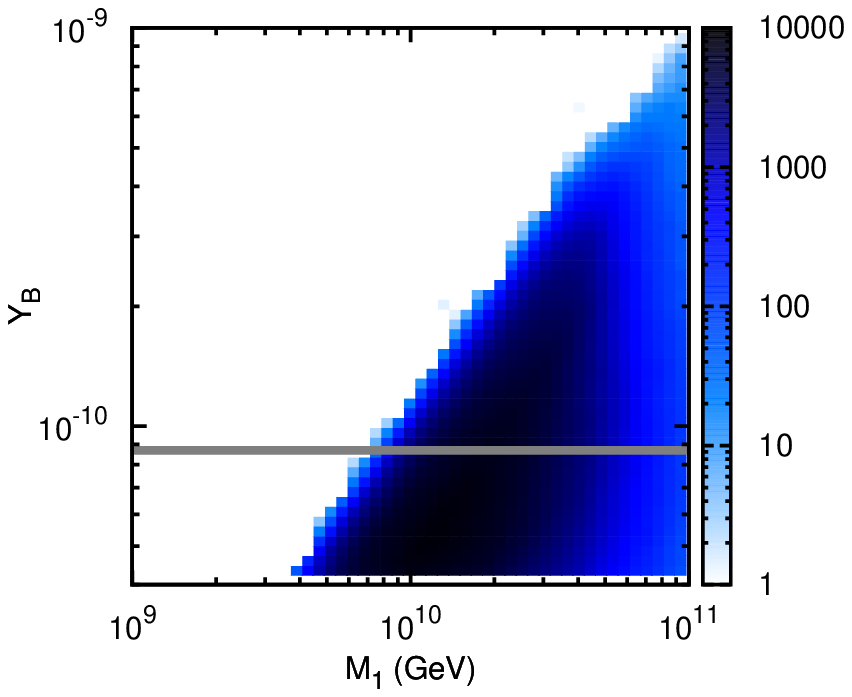}
    \includegraphics[width = 0.495\textwidth]{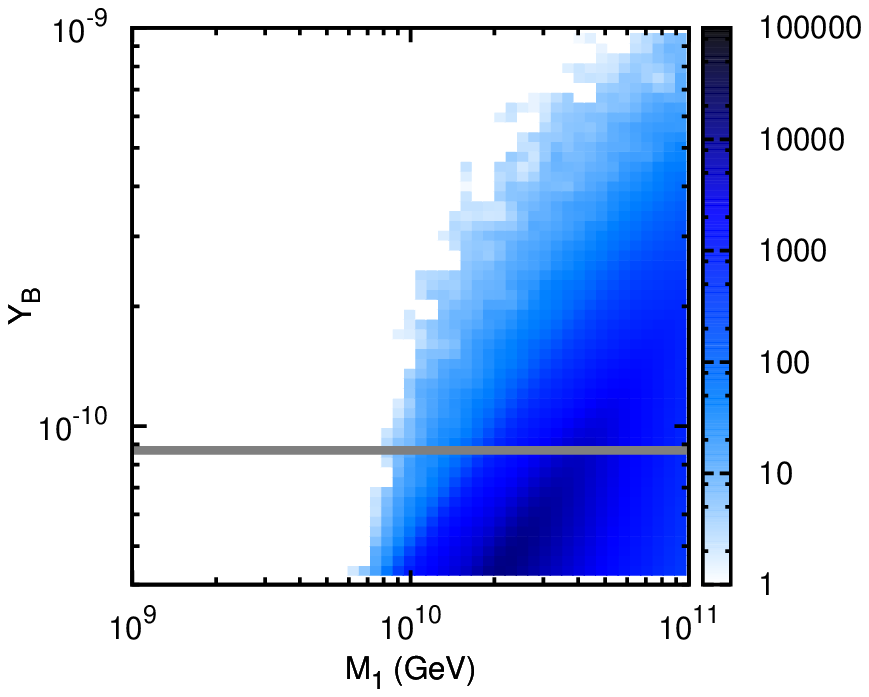}
  \caption{
Density of ``successful'' points, as a function of
 the baryon asymmetry and the lightest right-handed neutrino mass, assuming $\lambda_3 \sim 1$ and
$ \lambda_2/\lambda_1 \sim \lambda_3/\lambda_2 \sim 30$, for two different simulations: NH and $\tan \beta=50$ (left), and IH and $\tan \beta=2$ (right).
``Successful'' points are defined as for Figure \ref{Fig:M1_mt}.
}
  \label{Fig:bau_M1}
  \end{center}
\end{figure}

With the parametrisation described in section \ref{sample}, 
 the MCMC  easily  finds  larger  values of $M_1$ and $\tilde{m}$,
than the ``tuned''   points found analytically in Section \ref{Analytic}.
This preference for
larger $M_1$ is  expected, because the baryon asymmetry and 
right-handed neutrino masses are strongly correlated, 
see Fig. \ref{Fig:bau_M1} and eqn (\ref{eps}).
%
%


\begin{figure}[t!]
  \begin{center}
    \includegraphics[width = 0.495\textwidth]{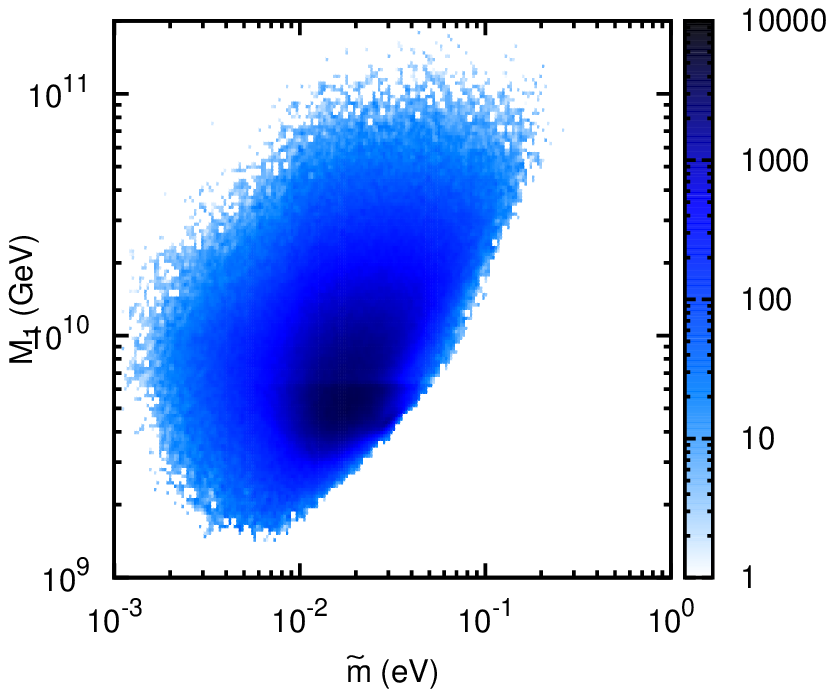}
    \includegraphics[width = 0.495\textwidth]{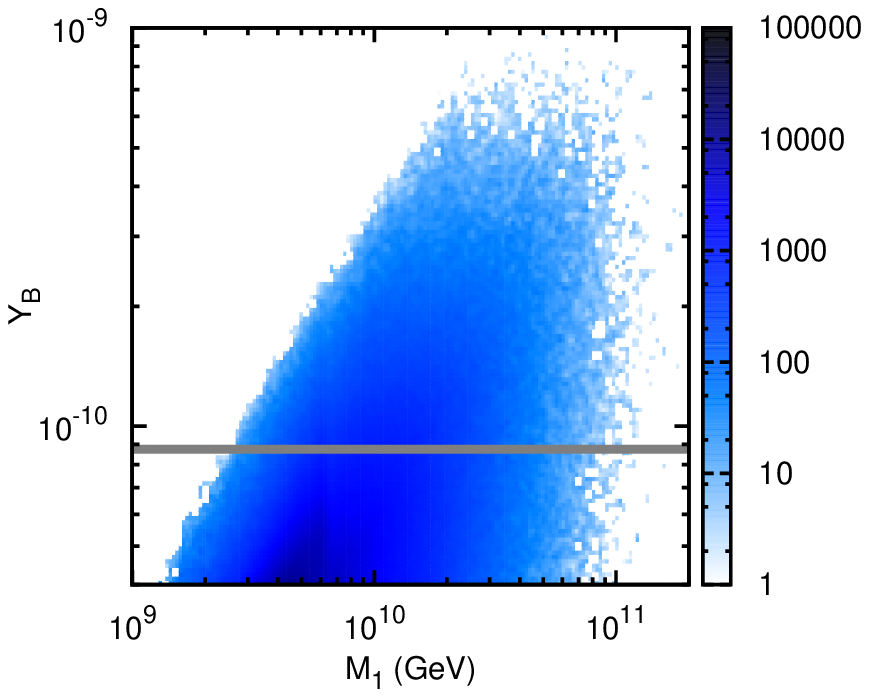}
  \caption{
Density of ``successful'' points, as a function of
the lightest right-handed neutrino mass $M_1$ and rescaled decay rate $\tilde{m}$, on the left-side, and between the baryon asymmetry and the lightest right-handed neutrino mass, on the right-side. We assume here $\lambda_3 \sim 1$ and $ \lambda_2/\lambda_1 \sim 100$ and $\lambda_3/\lambda_2 \sim 50$, for a NH in the light neutrinos and $\tan \beta=2$.
``Successful'' points are defined as for Figure \ref{Fig:M1_mt}.}
  \label{Fig:M1_mt_bau_M1steepY}
  \end{center}
\end{figure}

Nonetheless, as illustrated in Fig.\ref{Fig:M1_mt_bau_M1steepY},
 the MCMC succeeded in finding  points at lower $M_1$,
with a  steeper
\footnote{ The smallest yukawa must be small enough to 
ensure $\tilde{m} \sim m_*$.}
 hierarchy in the yukawas $\lambda_3 \sim 1$, 
$ \lambda_2/\lambda_1 \sim 100$ and $\lambda_3/\lambda_2 \sim 50$. 
The difficulties of finding
these tuned points are  discussed in section \ref{convergence}.

%

The importance 
of the $\sim  2$ decrease in $M_1$ and $\tilde{m}$,
at the tuned points,
is unclear to us: 
the cosmological
bound is on $T_{RH}$, rather than  $M_1$.  Since
in  strong washout,  an equilibrium population of 
$N_1$ can be generated  for  $T_{RH} \gsim M_1/5$,
  the points found
by the MCMC  at $M_1 \sim 10^{10}$ GeV, 
 could perhaps   generate the BAU at
the same $T_{RH}$ as the analytic points.   
In any case, we see in Fig.\ref{Fig:bau_M1} that the fraction of points with
big enough $Y_B$ is  very sensitive to $M_1$, and  therefore to
details of the complicated reheating/preheating process.

\begin{figure}[t!]
  \begin{center}
    \includegraphics[width = 0.495\textwidth]{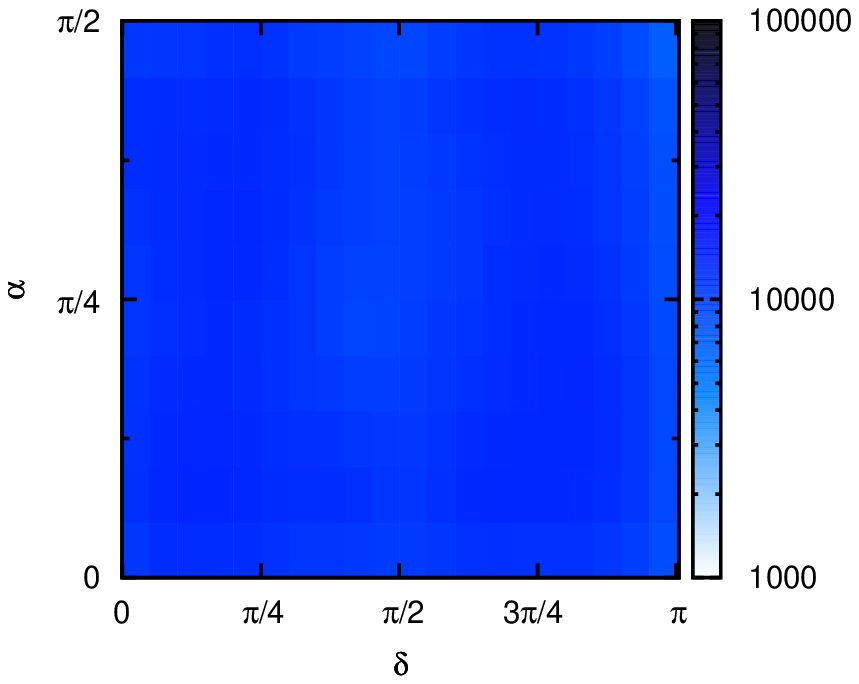}
    \includegraphics[width = 0.495\textwidth]{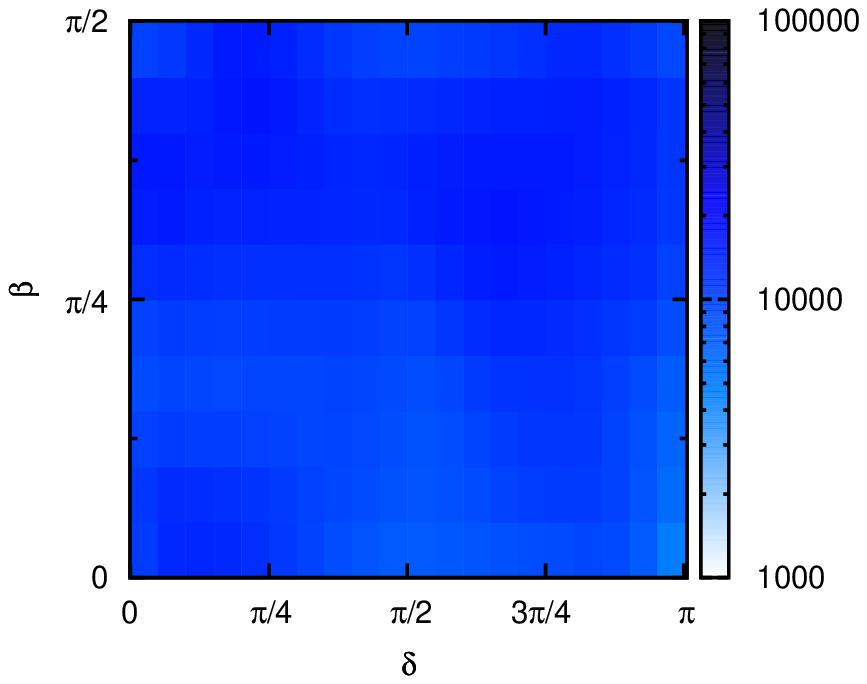}
    \includegraphics[width = 0.495\textwidth]{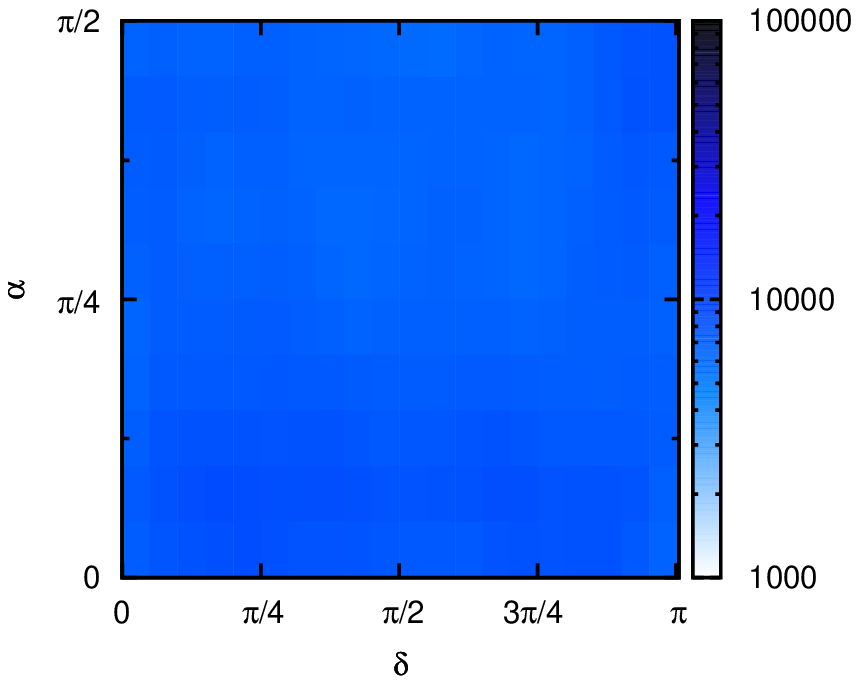}
    \includegraphics[width = 0.495\textwidth]{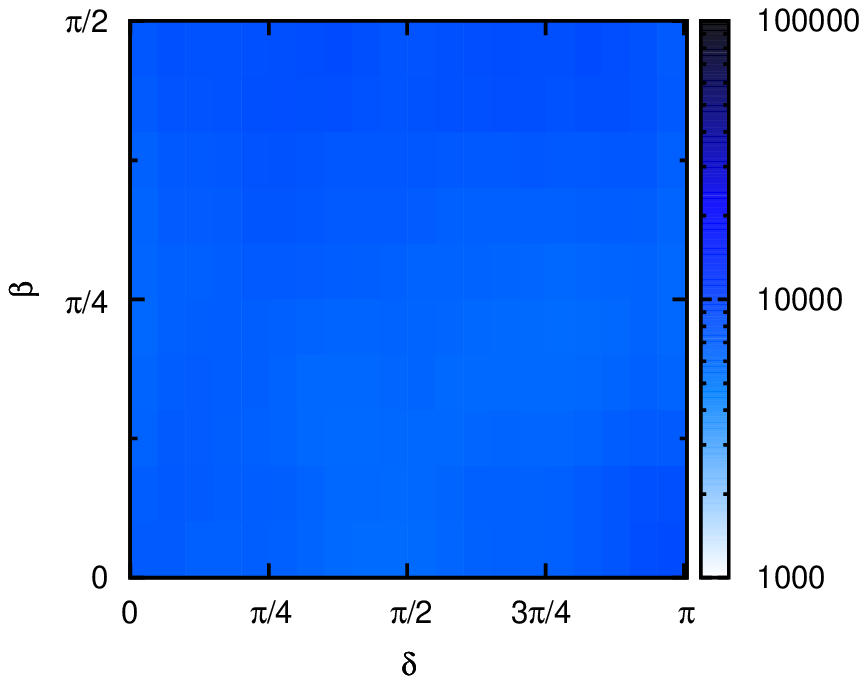}
    \caption{Density plots in the plane of the low energy phases 
$\delta-\alpha$ and $\delta-\beta$ in models with $\lambda_3 \sim 1$ and
$ \lambda_2/\lambda_1 \sim \lambda_3/\lambda_2 \sim 30$. Upper plots 
correspond to a simulation with NH and $\tan \beta=50$, and lower plots 
to IH and $\tan \beta=2$.
``Successful'' points are defined as for Figure \ref{Fig:M1_mt}.}
  \label{Fig:deltabeta}
  \end{center}
\end{figure}

\begin{figure}[t!]
  \begin{center}
    \includegraphics[width = 0.3\textwidth]{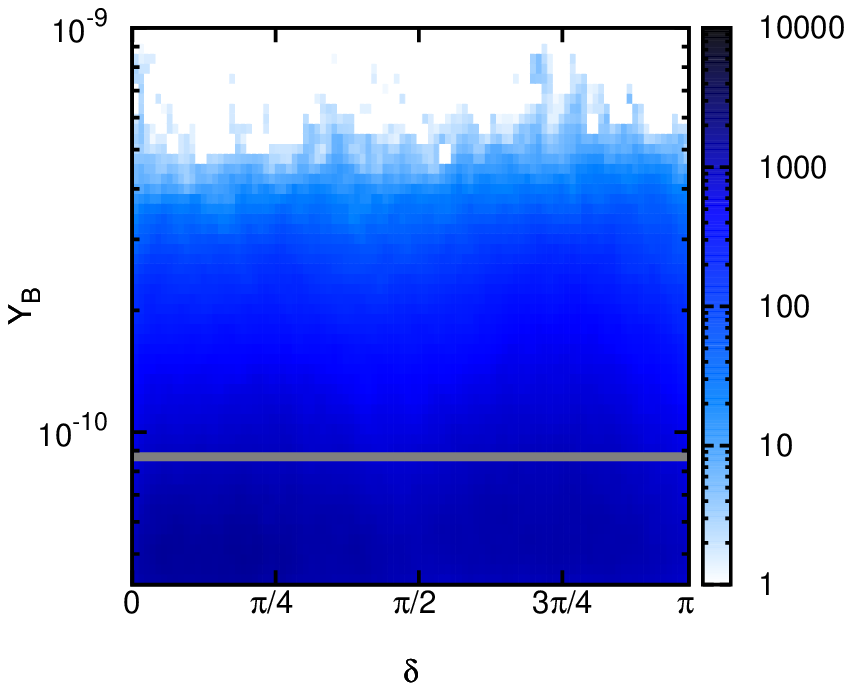}
    \includegraphics[width = 0.3\textwidth]{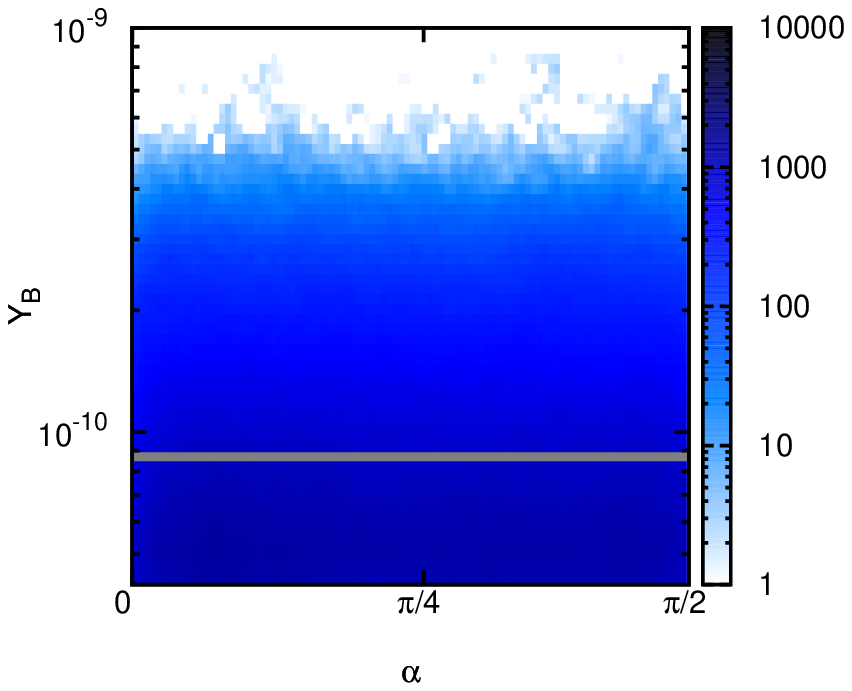}
    \includegraphics[width = 0.3\textwidth]{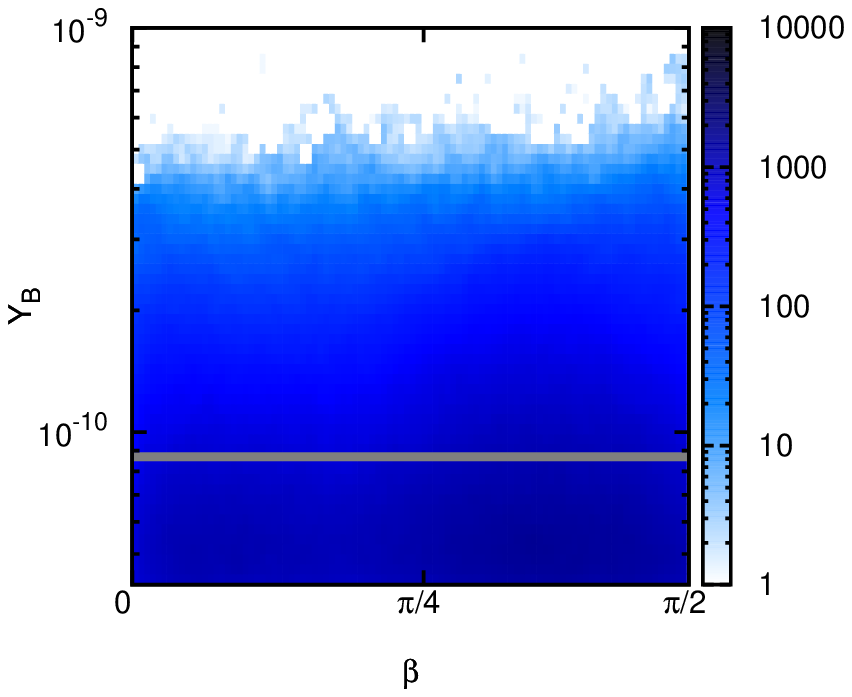}
    \includegraphics[width = 0.3\textwidth]{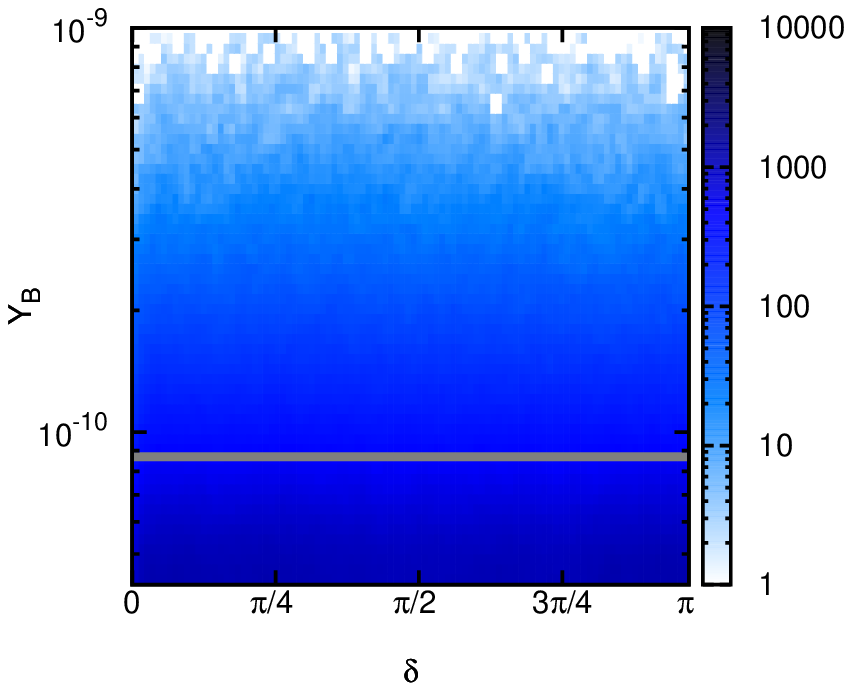}
    \includegraphics[width = 0.3\textwidth]{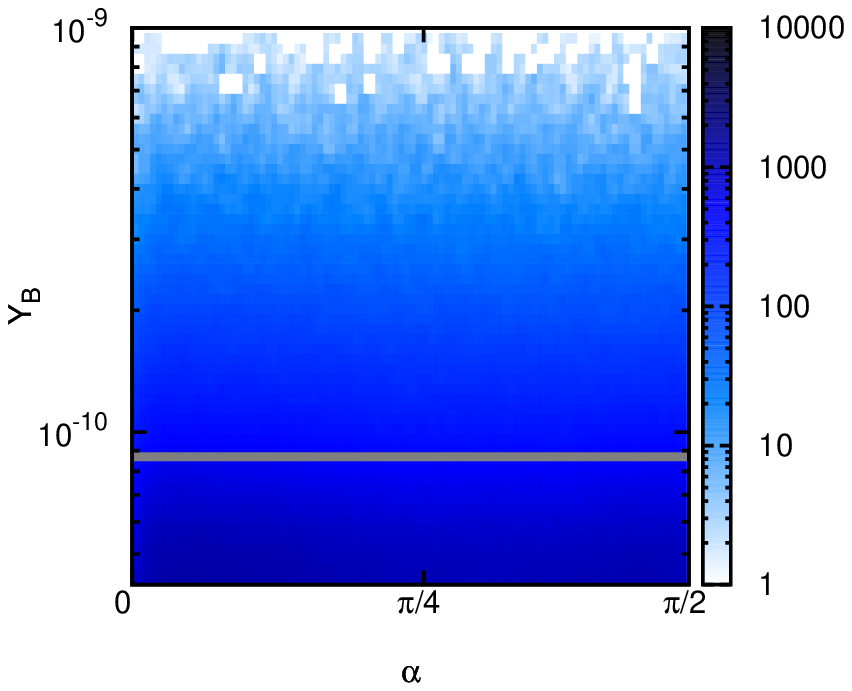}
    \includegraphics[width = 0.3\textwidth]{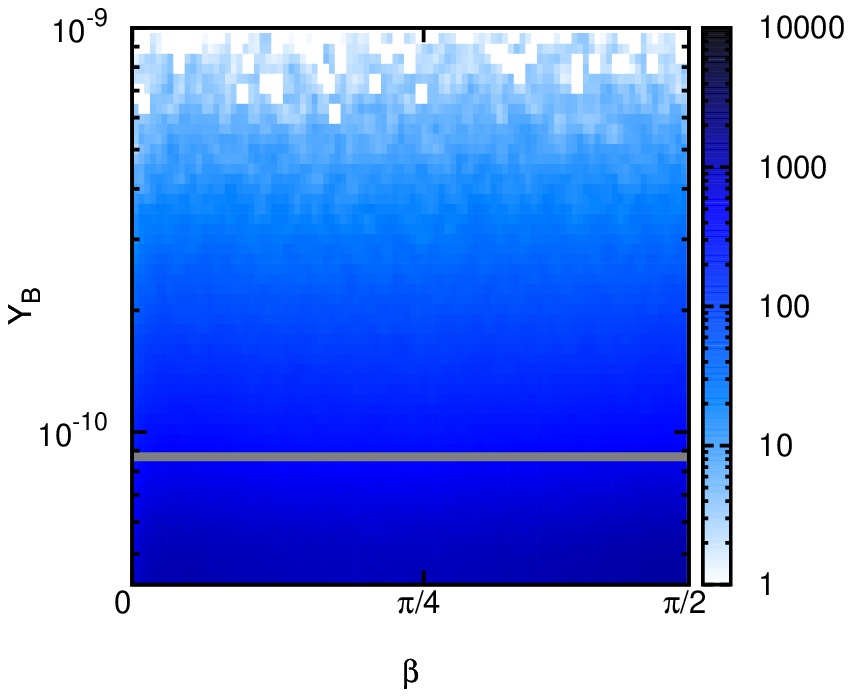}
    \caption{
Density of ``successful'' points, as a function of 
the BAU and the low energy phases in models with $\lambda_3 \sim 1$ and
$ \lambda_2/\lambda_1 \sim \lambda_3/\lambda_2 \sim 30$. Upper plots 
correspond to a simulation with NH and $\tan \beta=50$, and the lower plots 
to IH and $\tan \beta=2$.
``Successful'' points are defined as for Figure \ref{Fig:M1_mt}.}
  \label{Fig:bau_ph}
  \end{center}
\end{figure}

\begin{figure}[t!]
  \begin{center}
    \includegraphics[width = 0.495\textwidth]{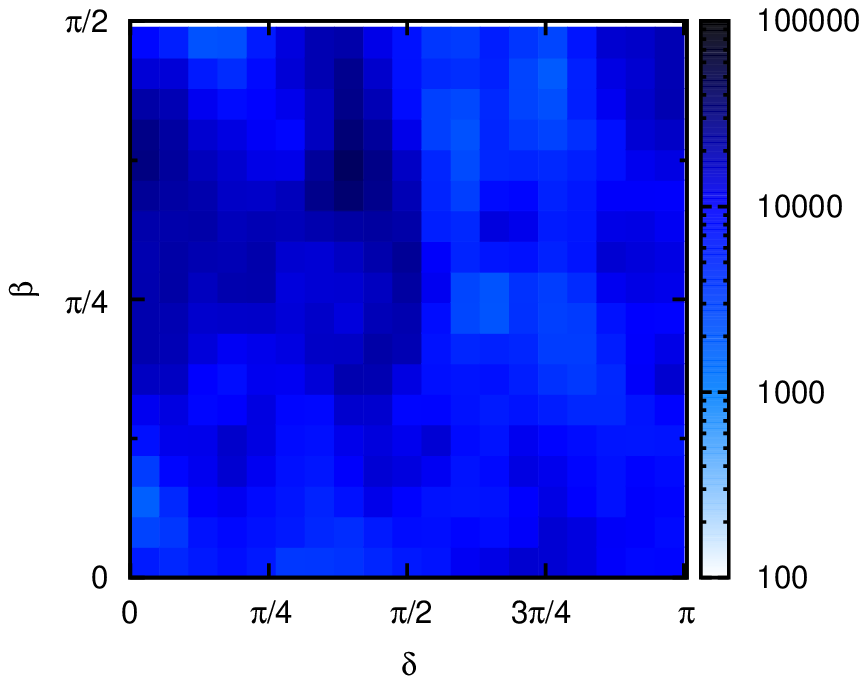}
    \includegraphics[width = 0.495\textwidth]{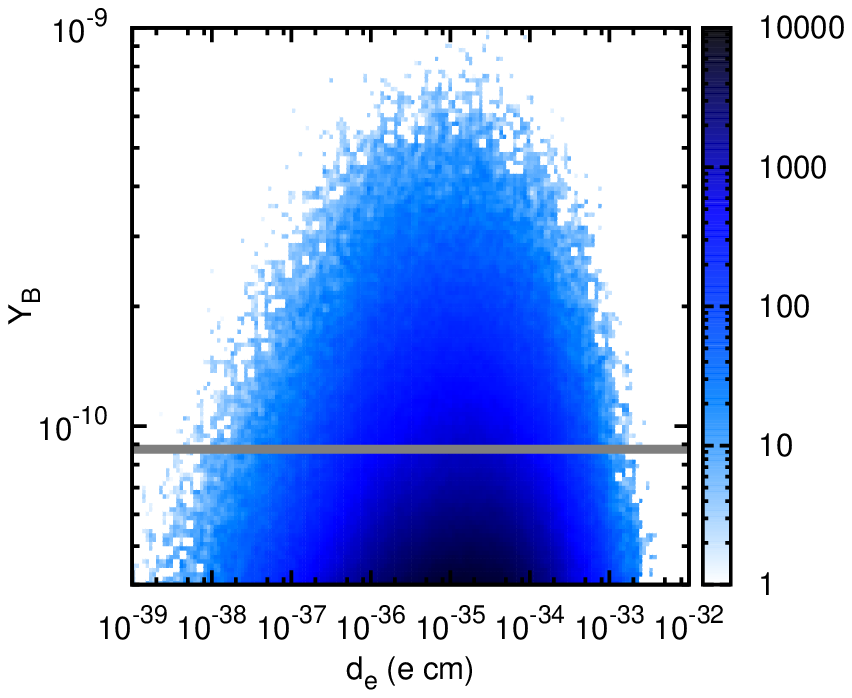}
  \caption{Density plot in $\delta-\beta$ plane and correlation between the baryon asymmetry and the electron EDM generated by neutrino yukawas. We assume here $\lambda_3 \sim 1$, $ \lambda_2/\lambda_1 \sim 100$ and $\lambda_3/\lambda_2 \sim 50$, for a NH in the light neutrinos and $\tan \beta=2$.
``Successful'' points are defined as for Figure \ref{Fig:M1_mt}.}
  \label{Fig:db_bau_edmsteepY}
  \end{center}
\end{figure}

In Fig. \ref{Fig:deltabeta}, we show density plots of the points resulting 
from our Markov Chains, corresponding to the 
the yukawa hierarchy $\lambda_2/\lambda_1 \sim \lambda_3/\lambda_2 \sim 30$, 
with $\lambda_3 \sim 1$, for 
normal hierarchy (NH) of the light neutrino masses 
and $\tan\beta =2$, and for 
inverse hierarchy (IH) and $\tan\beta =50$. 
In Fig. \ref{Fig:db_bau_edmsteepY} (plot on the left) we show a density plot in the $\delta-\beta$ plane for
$\tan \beta = 2$, NH and the steeper hierarchy  $ \lambda_2/\lambda_1 \sim 100$, $\lambda_3/\lambda_2 \sim 50$ and $\lambda_3 \sim 1$.
From those plots we  see that,
 for any value of the phases $\delta$, $\alpha$ and $\beta$
our conditions are satisfied. The analytic
results of Section \ref{Analytic}
agree with this. Thus, we can conclude that the baryon asymmetry of 
the universe  is 
{\it insensitive} to the low energy PMNS phases,
even in the ``best case'' where we see
MSUGRA-mediated lepton flavour violating processes.
 For completeness we also
show correlation plots between the generated BAU and the three low energy
phases in Fig.\ref{Fig:bau_ph}.
The low energy observables do not depend on $\tan \beta$, because we 
assume the $(g-2)_\mu$ discrepancy is due to slepton
loops, and we  use it to normalise
the LFV rates (see Eqn. \ref{BR_LFV}). On the contrary, the value of $\tan \beta$
is relevant in leptogenesis because it changes the number of distinguishable
flavours. However, as we can see comparing plots for small/large 
$\tan \beta$, the value of $\tan \beta$ does not change our conclusions.

\begin{figure}[t!]
  \begin{center}
    \includegraphics[width = 0.495\textwidth]{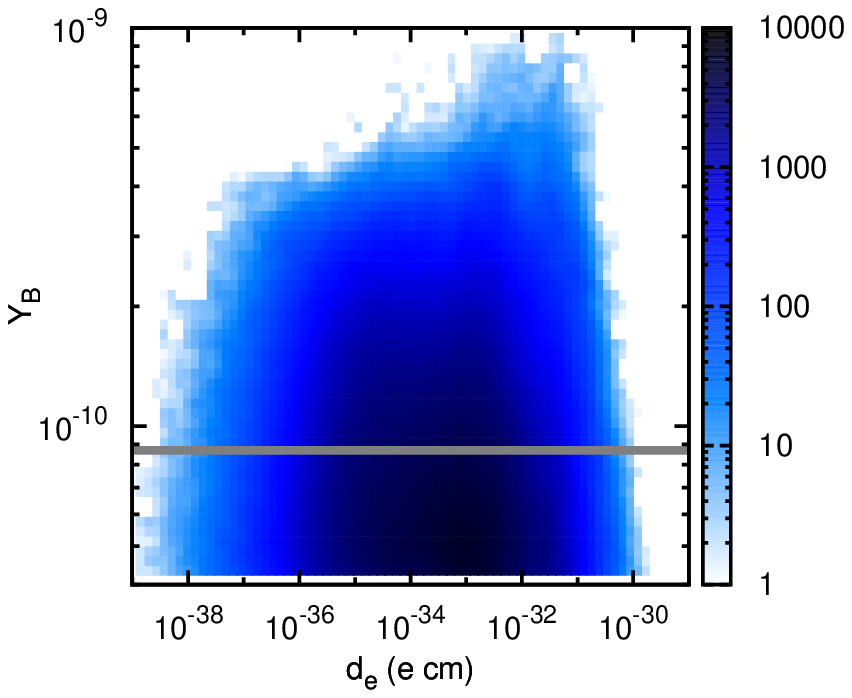}
    \includegraphics[width = 0.495\textwidth]{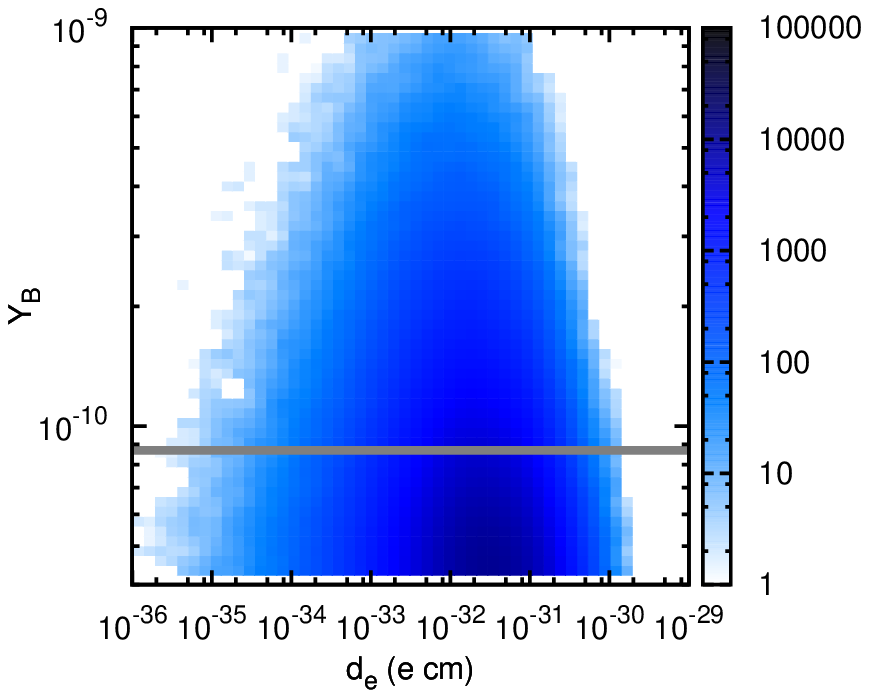}
    \caption{
{Density of ``successful'' points, as a function of
the baryon asymmetry and the electron EDM generated by neutrino yukawas in 
models with $\lambda_3 \sim 1$ and
$ \lambda_2/\lambda_1 \sim \lambda_3/\lambda_2 \sim 30$. The left panel  
corresponds to a simulation with NH  and $\tan \beta=50$, and the right panel 
to IH and $\tan \beta=2$.
``Successful'' points are defined as for Figure \ref{Fig:M1_mt}.}}
  \label{Fig:corr_de}
  \end{center}
\end{figure}

In Figs.\ref{Fig:corr_de} and \ref{Fig:db_bau_edmsteepY} (plot on the right), we plot the  contribution   to the
electric dipole moment of the electron, arising in
the MSUGRA seesaw with real soft parameters
at the high scale. 
For both   low and large $\tan\beta$, 
 points  from our MCMC  
generate an electron EDM $\lsim 10^{-30}e$cm.
 This agrees with the
results of \cite{Farzan:2004qu,Ellis:2001yza,JMR}.

%

\section{Summary}
\label{Conclusions}

The aim of this work was to study whether
the baryon asymmetry produced by thermal
leptogenesis was  sensitive to the
``low energy'' phases present in the
leptonic mixing matrix $U_{ PMNS}$.  
We considered  the 
 three generation type-I  supersymmetric seesaw model,
 in the framework of MSUGRA with real soft parameters
at the GUT scale, and  required that it
reproduces  low energy neutrino  oscillation data, 
 generates  a large enough 
baryon asymmetry of the
Universe via flavoured leptogenesis and 
 induces lepton flavour violating rates  within a few orders
of magnitude of current bounds.  We then enquired whether 
a preferred range 
for the low energy PMNS phases $\delta$ and $\beta$ can be predicted.

We used a ``left-handed'' bottom-up parametrisation
of the seesaw. Our parameter space scan was performed 
by a  Monte Carlo  Markov Chain  (MCMC), which
allows to efficiently explore high-dimensional spaces. 
It prefers to find the right-handed neutrino mass
$M_1 \gsim 10^{10}$ GeV, but can also
find successful points  with a smaller $M_1$ 
if it takes small steps in the 
relevant area of parameter space.
In this area, 
we can also show analytically that the
baryon asymmetry is insensitive to the
PMNS phases.

We have checked that there is no correlation between successful leptogenesis
and the low energy CP phases. That is: for any value of the low energy phases, 
the unmeasurable high energy parameters and the still unmeasured 
$m_1$ and $\theta_{13}$ 
can be arranged in order to have 
successful leptogenesis and LFV rates in the next round of experiments.
The analytic estimates indicate that this result will still be true even if 
$m_1$ and $\theta_{13}$ are measured and fixed to their experimental 
values.
Finally, we have estimated, for each point in our
chains,   the contribution of the 
complex neutrino Yukawa couplings to the electric
dipole moment of the electron. 
As expected, we find it to be $\lsim 10^{-30} e$cm,
just beyond the reach of next generation experiments.

\section*{Acknowledgments}
We would like to thank Yasaman Farzan, Filipe Joaquim, Martin Kunz, 
Isabella Masina, Miguel Nebot and 
Oscar Vives for useful discussions.  
J.G. is supported by a MEC-FPU Spanish grant.
This work is supported in part  by the Spanish grants 
FPA-2007-60323 and  FPA2005-01269, by the MEC-IN2P3 grant 
IN2P3-08-05 and by the EC RTN network MRTN-CT-2004-503369.

\appendix

\section{Fine tuning of the analytic points}
\label{Appft}

In this Appendix, we estimate the fine-tuning
of the points discussed in section \ref{Analytic},
with respect to the parametrisation  of
section \ref{Reconstructing}, which is used by the MCMC.

We do this in two steps.  
  First, in the parametrisation of
section \ref{Analytic},
we estimate the $3 \times 3$  matrix $W^\dagger =U^\dagger  V_L^\dagger  $
 which diagonalises $m$ in the basis where $\lambda\lambda^\dagger $
is diagonal. Approximating this diagonal  Yukawa basis 
to be the one  where 
$\hat{\Lambda}\hat{\Lambda}^\dagger$  is diagonal, we obtain:
\beq
W^\dagger = [\delta W]^\dagger \left[
\begin{array}{cc}
1 & ~0 ~~~0 \\
\begin{array}{c}
0 \\ 0 \\
\end{array}
& \hat{W}
\end{array} \right]
\eeq
where $ [\delta W]^\dagger$ is the small rotations that
rediagonalise $m = (\Delta_{ij} + \hat{D}_\kappa)v_u^2$,
and $\Delta_{ij} = \lambda_{i1} \lambda_{j1}/M_1$. If
$W^\dagger$ is parametrised as in eqn (\ref{V_L})
(but neglecting phases for simplicity), we find
\bea
\theta^W_{13} & \simeq & \frac{\Delta_{13}}{\kappa_3} \cos \hat{\theta}_W
+  \frac{\Delta_{12}}{\kappa_2} \sin \hat{\theta}_W \\
\theta^W_{12} & \simeq & - \frac{\Delta_{13}}{\kappa_3} \sin \hat{\theta}_W
+  \frac{\Delta_{12}}{\kappa_2} \cos \hat{\theta}_W \\
\sin \theta^W_{23} & \simeq & \sin \hat{\theta}_W 
+
\frac{\Delta_{23}}{\kappa_3} \cos \hat{\theta}_W ~~~.
\eea
To obtain $\lambda_{21}$ negligeable  compared to $\lambda_{31}$
in eqn (\ref{fin5.5}), requires no
particular  tuning of   $ \theta^W_{ 12}$ and $\theta^W_{ 13}$ with respect to  
$ \lambda_{21}$ and$ \lambda_{31}$.

The second step is to estimate the tuning required
to  obtain small angles $\theta^W_{12}$ and $\theta^W_{13}$
in $W^\dagger = U^\dagger V_L^\dagger$.  With
$V_L^\dagger$ parametrised as in eqn (\ref{V_L}),
this happens if   the angles of $V_L$  satisfy 
$\theta^L_{ ij}  \simeq \theta_{ij}$
(for $i,j = 12, 13$). So the ``tuning'' required 
in  $\theta^L_{ 12} $ and $\theta^L_{ 13} $   to obtain
small  $\theta^W_{ ij}  = \theta^L_{ ij} -  \theta_{ ij} $ is 
\beq
\frac{\theta^W_{12}}{\theta^L_{12}}
\frac{\theta^W_{13}}{\theta^L_{13}}
\simeq 
\frac{\tilde{m}^2}{m_3^2\theta_{13}}   
\eeq
This implies that $\theta_{13}^L$ 
must be tuned against  $\theta_{13} $
to obtain  $\theta^W_{13} \sim .01$.
 If instead
 $\theta_{13} \lsim .01$, there is no particular tuning
of $\theta^W_{13}$, and the tuning of 
  $\theta^W_{12}$ with respect to  $\theta^L_{12}$
is or order $\tilde{m}/m_3$.

\end{document}